%% file: main.tex
\documentclass[12pt]{article}
\include{prembles}

\usepackage{authblk}
\usepackage[left=.71in, right=.71in,top=.7in,bottom=.7in]{geometry}

\usepackage[normalem]{ulem}
\usepackage{xcolor}
\definecolor{forestgreen}{rgb}{0.33,0.61,0.34}

\usepackage[normalem]{ulem}
\usepackage{xcolor} 

\usepackage{url}

\begin{document}

\title{I2E2S2R Rumor Spreading Model in Homogeneous Network with Hesitating and Forgetting Mechanisms}

	\author[1]{\small Md. Nahid Hasan  }
	\author[2]{\small Sujana Azmi Polin }
    \author[2,3]{\small Saiful Islam }
    \author[2]{\small Chandra Nath Podder \thanks{Corresponding author.  {\it E-mail address:} cpodder@du.ac.bd}}
    
   	\affil[1]{\it\footnotesize Department of Mathematics, Bangladesh University of Engineering and Technology, Dhaka 1000, Bangladesh}
   	\affil[2]{\it\footnotesize Department of Mathematics, University of Dhaka, Dhaka 1000, Bangladesh}
     \affil[3]{\it\footnotesize Institute for Artificial Intelligence and Data Science, University at Buffalo, State University of New York at Buffalo, NY, USA}
   
   	\date{}
\maketitle

	\noindent\rule{170mm}{0.25mm} 
\section*{Abstract}
The spreading and controlling of rumors have great impacts on our society. The transmission of infectious diseases and the spreading of rumors have some common scenarios. Like cross-infection propagation of diseases, two or many kinds of rumors or information may spread at the same time. 
In this paper, we propose a novel I2E2S2R rumor-spreading model in a homogeneous network. The rumor-free equilibrium, as well as the basic reproduction number, have been calculated from the mean-field equations of the model. Lyapunov function and the LaSalle invariance principle are used to establish the global stability of the rumor-free equilibrium. In numerical simulations, it is perceived that a higher degree of network helps to spread rumors quickly. We have also found that making people aware can help to disappear rumors faster from the network. In addition, making people divert from the rumor to exact information can lessen the spreading of the rumor.

\noindent \textbf{Keywords}\\
\noindent Double-rumors Spreading, Epidemic Model, Multiple rumors model, Basic Reproduction Number, Stability, Homogeneous Network\\
\noindent\rule{170mm}{0.25mm} \\
	
\section{Motivation}
Rumour is usually defined as the unverified elaboration or annotation of the public interesting things, events, or issues
that circulate via multiple channels, in itself neither true nor false \cite{GALAM2003571,hayakawa2002sociology,kimmel2004rumors, KOSFELD2005646}. Rumour is a form of social communication and can
shape public opinion and affect the beliefs of
individuals, which can lead to the changes of
individual's attitude towards economic, political,
and social aspects \cite{kawachi2008deterministic,misra2012simple}. 
Rumors are part of our everyday life. As an important part of people's lives, rumors are being used as a special weapon of public opinion and can potentially impact social life \cite{daley1964epidemics}. As an aftermath, it leads to social panic and instability \cite{barabasi1999emergence,kosfeld2005rumours,kesten2005spread}. With the growing presence of online social networks, rumors are no longer disseminated by word of mouth over a small area. However, they are
disseminated amongst strangers in different regions and different countries of the world. Rumors travel faster and more extensively over the internet than by conventional means.\\

Rumors that trigger strong emotions such as fear, outrage, or curiosity have a higher chance of going viral on the internet. Social media algorithms tend to favor sensationalized or inaccurate content that receives much interaction. Unchecked rumors can harm relationships and undermine trust in institutions. For quite some time, rumors surrounding public health and health emergencies have been causing panic and hysteria among people. Falsehoods spread by rumors create divisions in societies by promoting mistrust between different groups. To promote unity and cooperation, communities must learn to control rumors. Controlling rumors can prevent unnecessary fear and chaos. Traditional approaches to managing rumors, such as fact-checking, debunking, or censorship, are reactive and often have limited effectiveness since they address rumors only after they have already spread widely. In contrast, a proactive and systematic framework, such as an optimal control approach, can be used strategically.
\par

Epidemic spreading and rumor propagation have similar propagation mechanisms. That is why epidemic models have been widely used to investigate the rumor propagation dynamics of online social networks. The study of rumor-spreading models began in the 1940s. Based on the SIR epidemic model, Daley and Kendall \cite{daley1964epidemics,daley1965stochastic} proposed the basic DK model, the beginning of rumor-spreading modeling, in the 1960s. In their
model, the population is partitioned into three groups: those who are unaware of the rumor (ignorants), those who spread the rumor (spreaders), and those who are aware of the rumor but choose not to spread it (stiflers). After that, Maki \cite{maki1973mathematical} modified the DK model and developed the MK model, in which rumors propagate through direct contact between spreaders and others. Afterward, Nekovee et al. \cite{nekovee2007theory} and Isham et al. \cite{isham2010stochastic} built a new model by combining the MK model with the SIR epidemic model on complex networks. Gu et al. \cite{gu2007forget,gu2008effect} and Zhao et al.
\cite{zhao2012sihr,zhao2013rumor} refined the rumor-spreading model by incorporating the effects of the remembering mechanism in complex networks. Wang et al. \cite{wang20142si2r} developed a 2SI2R rumor-spreading model to investigate the behavior of two rumors spreading simultaneously, and Zhang and Zhu \cite{zhang2018stability} extend this model for the complex network. Liu et al. \cite{liu2018rumor} develop an SEIR rumor-spreading model with a hesitating mechanism, where only a single rumor spreading is considered. Lately, Hasan et al. \cite{hasan2021ivesr} used the idea of vaccination in the rumor-spreading model to investigate the behavior of rumor spreading and to reduce the spreading of rumors. \par

It's essential to understand the gravity of rumor-mongering and the detrimental effects it can have on our lives. In order to avoid these negative consequences in various aspects of our daily routine, it is imperative to put a stop to the spreading of any such rumors. Optimal control is the key to achieving the desired outcomes for any given system while adhering to predefined constraints. In the case of rumor propagation models, the goal is to reduce the number of people infected by the rumor while keeping control costs to a minimum. Researchers have proposed several promising approaches to achieve this objective that have yielded positive results.  Zhao et al. \cite{zhao2014dynamical}  found that the best way to control the spread of rumors is through a combination of government-released information and network monitoring. Wang et al., in their SIS network model, have shed light on the robust role media coverage can play in the spread of diseases \cite{wang2013impact}. The study revealed that media coverage can significantly impact individual behavior towards rumors, making it a crucial factor to consider in any strategy to control the spread of disease. Furthermore, scholars have developed a theoretical framework that analyzes the participatory social reporting phenomenon and its influence on the spread of rumors \cite{oh2013community}. Cost management, feedback mechanisms, super-spreading events, multiplex network interactions, and competitive information propagation are all taken into account in a number of novel models \cite{jain2023optimal,yao2019research,tian2015ssic,zhang2019interacting,yang2020containment}.

All the existing works about two rumors spreading simultaneously are based on the SIR model. In the SIR model, there is no exposed class, which is equivalent to the hesitating compartment. So, using the SEIR model in this regard is more reasonable. Usually, when a person hears a rumor, they do not start spreading the rumor instantly. The person first thinks about the rumor, whether it is true or false, beneficial for them or not, etc. Some people assume the rumor is false at first glance and never spread the rumor. In addition, in the existing model \cite{wang20142si2r,zhang2018stability,zan2018dsir}, it is assumed that when a spreader forgets a rumor, it turns into a stifler. In reality, after forgetting the rumor, when the person comes in contact with a spreader and recalls the rumor, they may start spreading the rumor. So, it is more convenient to consider that the spreader becomes a stifler after forgetting the rumor. However, these scenarios can not be described meaningfully using the SIR model.

In today's fast-paced world, social media has become an integral part of our lives. However, with the increasing amount of information circulating on the network, it's crucial to identify the authenticity of the news. To prevent false rumors from spreading like wildfire, it's essential to inform people of the actual events. While traditional methods of spreading awareness are suitable, using a combination of strategies can be more effective. Social media influencers can play a critical role in raising awareness by using their platforms to spread the right message. We understand that traditional media like TV and radio are still relevant and effective, which is why we incorporate these strategies to control the spread of rumors optimally. By using a mix of traditional and modern techniques, we can combat false information and ensure that the truth prevails.

In this paper, we develop an I2E2S2R based on the SEIR model for two information spreads simultaneously. Either the two pieces of information can be rumors or one of them is a rumor, and the other is the corresponding accurate information of the rumor.


\section{Model formulation}
We assume that the entire population forms a social network. Additionally, we consider that there are types of information: either both are false, or one is false. At the same time, the other provides accurate details on the same subject, propagated within the networks.
We refer to these two types of information as Information 1 and Information 2. We divide the entire population into seven distinct classes: Ignorant($I$), Exposed 1 ($E_1$),  Exposed 2 ($E_2$), Spreader 1 ($S_1$), Spreader 2 ($S_2$), Stifler 1 ($R_1$), Stifler 2 ($R_2$). Ignorant consists of individuals who have had no prior acquaintance with the rumors and are susceptible to acquiring knowledge about the rumors. Individuals in the classes Exposed 1 and Exposed 2 are those who have been informed of Rumor 1 and Rumor 2 or exact information corresponding to Rumor 1, respectively. However, they are in a hesitating state about spreading the rumor. The individuals actively spreading out information 1 and information 2, respectively, are in the classes $S_1$ and $S_2$. $R1$ and $R_2$ denote those individuals who have known information 1 and information 2, respectively, but will not spread it anymore. During the propagation of the rumor, individuals enter and depart an area; some individuals do not use social networks but can create new accounts and become active in the network. While some individuals are active on social networks, they may deactivate their accounts. As a result, we assume the population is open, and the dynamics include both the processes of entering and departing. Our model considers the newly entering individuals to be classified as ignorant.

 The following rules govern the rumour-spreading procedure among seven compartments. We consider a homogeneous network with average degree $\overline{k}$. If a spreader individual in $S1$ contacts an ignorant individual of the class $I$, then the ignorant transfers to the $E_1$ compartment with a probability $\lambda_1$. Similarly, after contact with a member of $S_2$, an ignorant transfer to $E_2$ compartment with a probability $\lambda_2$. That is, through contact with the spreaders, the ignorant individuals come to know about the rumor, and they enter into a hesitating period during which they contemplate whether to spread the rumor. Following the interaction of $I$ with $S_1$, a few portions of ignorant individuals do not trust the rumor, and they transfer to stifler compartment $R_1$ at a rate $\beta_1$ and for spreader 2, with the same distrust some of the ignorant transfer to stifler compartment at a rate $\beta_2$. Then, after hesitating, a few individuals from the $E_1$ and $E_2$ compartments decided to spread information 1 (rumor 1) and information 2 (rumor 2 or exact information corresponding to rumor 1) and turn into  $S_1$ and $S_2$ at rates $\tau_1$ and $\tau_2$ respectively. Also, after hesitating with self-realization, a few individuals from the $E_1$ and $E_2$ compartments decided not to spread either information and transferred into $R_1$ and $R_2$ at rates $\mu_1$ and $\mu_2$ respectively. Moreover, a few individuals in the $E_1$ compartment transferred into the $E_2$ with a probability $\sigma$ due to their interaction with $S_2$. With this contact, these individuals are influenced by the second information and prioritize it over the first. Later, from  $S_1$ and $S_2$ compartments, a portion of individuals transfer to $R_1$ and $R_2$ compartments with probability $\alpha_1$ and $\alpha_2$ respectively, due to the contacts with $R_1$ and $R_2$. Also, due to forgetting the rumor, spreader one and spreader 2 transfer to the ignorant compartment at rates $\eta_1$ and $\eta_2$, respectively. In addition, due to the contacts with spreader 2, a few individuals from spreader 1 and stifler 1 transfer to exposed 2 at probabilities $\gamma$ and $\psi$, respectively. We also assume that information 1 starts spreading first and information 2 starts spreading after a certain amount of time of spreading information 1. Additionally, we consider a constant entering rate of ignorant individuals denoted by $\xi$ and a constant leaving rate of individuals from any compartment, $\delta$. The I2E2S2R rumor-spreading procedure is shown in Fig.~\ref{modeldiag}  

Considering all the above assumptions, we obtain the following mean-field equations. See Table \ref{tbl1} for a description of the model parameters.

\begin{equation}\label{eqn1}
\begin{split}
    & \frac{dI(t)}{dt}=\xi+\eta_1S_1+\eta_2S_2-\overline{k}\lambda_1IS_1-\overline{k}\lambda_2IS_2-\overline{k}\beta_1IS_1-\overline{k}\beta_2IS_2-\delta I\\[5pt]
    & \frac{dE_1(t)}{dt}=\overline{k}\lambda_1IS_1-\tau_1E_1-\overline{k}\sigma E_1S_2-\mu_1E_1-\delta E_1\\[5pt]
    & \frac{dE_2(t)}{dt}=\overline{k}\lambda_2IS_2+\overline{k}\sigma E_1S_2+\overline{k}\gamma S_1S_2+\overline{k}\psi S_2R_1-\tau_2E_2-\mu_2E_2-\delta E_2\\[5pt]
    &\frac{dS_1(t)}{dt}=\tau_1 E_1-\eta_1 S_1-\overline{k}\alpha_1S_1R_1-\overline{k}\gamma S_1S_2-\delta S_1\\[5pt]
    &\frac{dS_2(t)}{dt}=\tau_2 E_2-\eta_2 S_2-\overline{k}\alpha_2S_2R_2-\delta S_2\\[5pt]
    &\frac{dR_1(t)}{dt}=\mu_1E_1+\overline{k}\beta_1IS_1+\overline{k}\alpha_1S_1R_1-\overline{k}\psi S_2R_1-\delta R_1\\[5pt]
    &\frac{dR_2(t)}{dt}=\mu_2E_2+\overline{k}\beta_2IS_2+\overline{k}\alpha_2S_2R_2-\delta R_2\\[5pt]
\end{split}
\end{equation}

\tikzstyle{hblock} = [rectangle,text width=6em,text centered,rounded corners]
\tikzstyle{block} = [rectangle,draw,fill=blue!20,text width=6em,text centered,rounded corners]
\tikzstyle{line} = [draw, -latex']

\begin{figure}[H]
\centering

\begin{tikzpicture}
\node [hblock,node distance=2cm] (h1) {   };

\node [hblock,below of=h1,node distance=2cm] (h2) {   };
\node [hblock,right of=h2,node distance=2cm] (h3) {   };
\node [block,right of=h3,node distance=4cm] (E1) {Exposed 1\\$E_1(t)$};
\node [block,right of=E1,node distance=4cm] (S1) {Spreader 1\\$S_1(t)$};
\node [block,right of=S1,node distance=4cm] (R1) {Stifler 1\\$R_1(t)$};
\node [block,below of=h3,node distance=2cm] (I) {Ignorant\\$I(t)$};
\node [hblock,below of=I,node distance=2cm] (h4) {   };
\node [block,right of=h4,node distance=4cm] (E2) {Exposed 2\\$E_2(t)$};
\node [block,right of=E2,node distance=4cm] (S2) {Spreader 2\\$S_2(t)$};
\node [block,right of=S2,node distance=4cm] (R2) {Stifler 2\\$R_2(t)$};
\node [hblock,left of=I,node distance=1cm] (h5) {   };

\path [line] (h5) --node[above]{$\xi$} (I);
\path [line] (I) --node[above]{$\lambda_1$} (E1);
\path [line] (I) --node[below]{$\lambda_2$} (E2);
\path [line] (E1) --node[above]{$\tau_1$} (S1);

\draw[->] (1.3,-4.5)--(1.3,-5.3) node[below](){$\delta I$};

\draw [->](9.7,-1.5) --(9.7,-1) node() {} --(2.3,-1) node [at end,above] () {$\eta_1$}--(2.3,-3.5) node () {};

\path [line] (S1) --node[above]{$\alpha_1$} (R1);

\draw [->] (E1.north) --(6,-.5) node() {} --(14,-.5) node [midway,above] () {$\mu_1$}--(14,-1.5) node () {};

\draw [->] (5.5,-1.5) --(5.5,-.5) node[above] () {$\delta E_1$};

\draw [->] (10.3,-1.5) --(10.3,-.8) node[midway,right] () {$\delta S_1$};

\draw [->] (R1.east) --(16,-2) node[right] () {$\delta R_1$};

\path [line] (E2) --node[below]{$\tau_2$} (S2);
\path [line] (S2) --node[below]{$\alpha_2$} (R2);
\path [line] (E1) --node[right]{$\sigma$} (E2);
\path [line] (S1) --node[right]{$\gamma$} (E2);
\path [line] (R1) --node[right]{$\psi$} (E2);

\draw [->](9.7,-6.5) --(9.7,-7) node() {} --(2.3,-7) node [at end,below] () {$\eta_2$}--(2.3,-4.5) node () {};

\draw [->] (E2.south) --(6,-7.5) node() {} --(14,-7.5) node [midway,below] () {$\mu_2$}--(14,-6.5) node () {};

\draw [->] (5.5,-6.5) --(5.5,-7.5) node[below] () {$\delta E_2$};

\draw [->] (10.3,-6.5) --(10.3,-7.3) node[midway,right] () {$\delta S_2$};

\draw [->] (R2.east) --(16,-6) node[right] () {$\delta R_2$};
\draw [->](1.8,-4.5) --(1.8,-8.2) node() {} --(14.5,-8.2) node [midway,below] () {$\beta_2$}--(14.5,-6.5) node () {};
\draw [->](1.8,-3.5) --(1.8,.3) node() {} --(14.5,.3) node [midway,above] () {$\beta_1$}--(14.5,-1.5) node () {};

\end{tikzpicture}

\caption{The schematic diagram of the I2E2S2R model for the rumor spreading process.}
\label{modeldiag}
\end{figure}
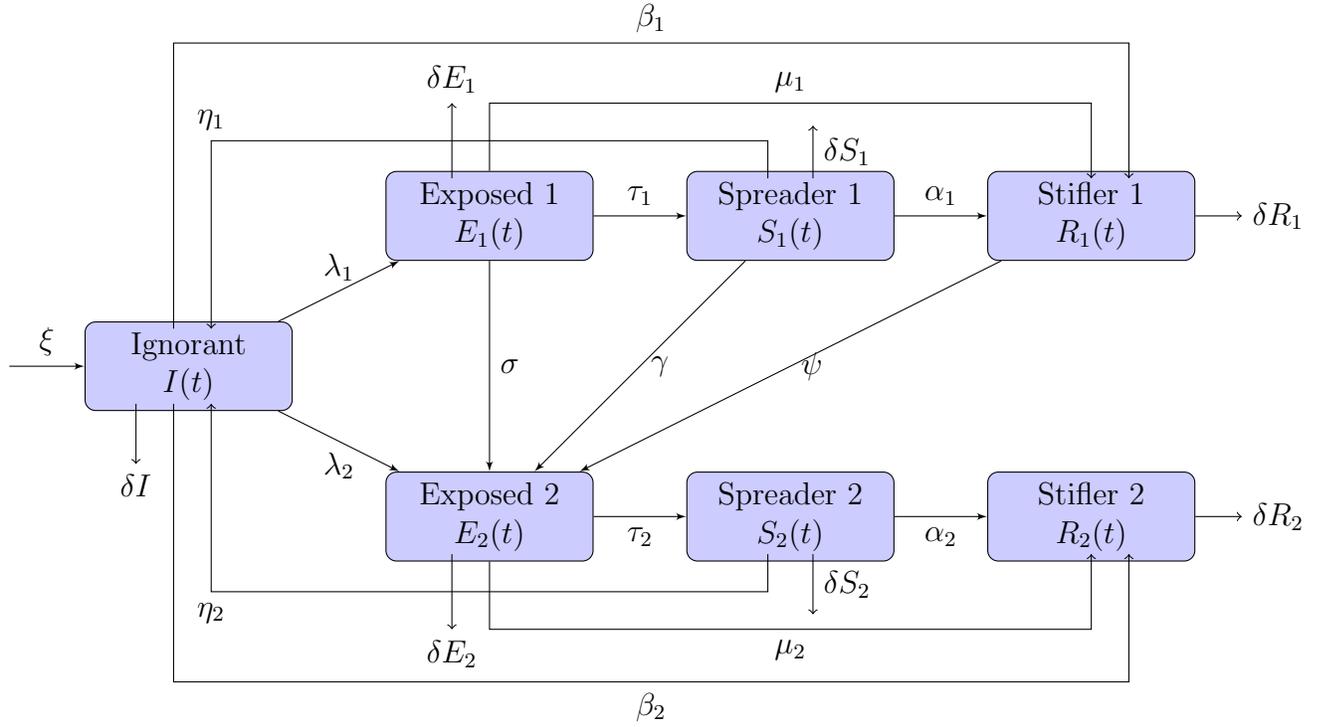

\begin{table}[H]
\centering
\rowcolors{3}{}{lightgray}

\begin{tabular}{ |c|l| }
\hline
\multicolumn{2}{|c|}{Parameter List} \\
\hline
Parameters & Explanation \\
\hline
$\xi$ & Entering rate of ignorants \\
$\delta$ & Leaving rate of the population  \\
$\lambda_1$ & Transmission probability from ignorant to exposed 1 \\
$\lambda_2$  & Transmission probability from ignorant to exposed 2 \\
$\beta_1$ & Rate at which the ignorant distrust the rumor 1\\
$\beta_2$ & Rate at which the ignorant distrust the rumor 2\\
$\tau_1$ & Transfer rate from exposed 1 to spreader 1\\
$\tau_2$ & Transfer rate from expose 2 to spreader 2\\
$\mu_1$ & Transfer rate from exposed 1 to stifler 1\\
$\mu_2$ & Transfer rate from exposed 2 to stifler 2\\
$\eta_1$ & Rate at which spreader 1 forgets the rumor 1\\
$\eta_2$ & Rate at which spreader 2 forgets the rumor 2\\
$\alpha_1$ & Transmission probability from spreader 1 to stifler 1\\
$\alpha_2$ & Transmission probability from spreader 2 to stifler 2\\
$\sigma$ & Transmission probability from exposed 1 to exposed 2 due to contacts with spreader 2\\
$\gamma$ & Transmission probability from spreader 1 to exposed 2 due to contacts with spreader 2\\
$\psi$ & Transmission probability from stifler 1 to exposed 2 due to contacts with spreader 2\\
\hline
\end{tabular}
\caption{The description of the parameters used in the model (\ref{eqn1}).}
\label{tbl1}
\end{table}
The population density of Ignorants,  Exposed 1,  Exposed 2, Spreader 1, Spreader 2, Stifler 1, Stifler 2  at any time $t$ are denoted by $I(t)$, $E_1(t)$, $E_2(t)$, $S_1(t)$, $S_2(t)$, $R_1(t)$ and $R_2(t)$ respectively, and the total number of individuals is $N(t)$. We have $I+E_1+E_2+S_1+S_2+R_1+R_2=1$ in normalized form. We also assume that the entering rate and leaving rate of individuals are equal, that is, $\xi =\delta$.

\section{Model analysis}
In this section, we compute the basic reproduction number, determine the equilibrium states, and investigate their stability properties.
\subsection{Rumor free equilibrium}

For the rumor-free equilibrium, we consider the count of the compartment of spreaders individuals $S^0_1 = 0$ and $S^0_2 = 0$. Then, we have
$
    I^0= \frac{\xi}{\delta}=1, \, 
    E^0_1= 0,  \, 
    E^0_2= 0, \, 
    S^0_1= 0, \,
    S^0_2= 0, \, 
    R^0_1= 0, \,
    R^0_2= 0.
$
Therefore, the rumor-free equilibrium is
\begin{equation}\label{eqn2}
    \overline{E}_0=\left( 1,0,0,0,0,0,0 \right)
\end{equation}
We explore the stability properties of the $\overline{E}_0$ of the following three subsections.
\subsection{Basic reproduction number ($\mathcal{R}_0$) }

We calculate the basic reproduction number of the
model (\ref{eqn1}) using next generation matrix method as in \cite{van2002reproduction}. Consider the compartments that are related to the spreader to obtain the following subsystem.
\begin{equation}\label{eqn3}
\begin{split}
    & \frac{dE_1(t)}{dt}=\overline{k}\lambda_1IS_1-\tau_1E_1-\overline{k}\sigma E_1S_2-\mu_1E_1-\delta E_1\\[5pt]
    & \frac{dE_2(t)}{dt}=\overline{k}\lambda_2IS_2+\overline{k}\sigma E_1S_2+\overline{k}\gamma S_1S_2+\overline{k}\psi S_2R_1-\tau_2E_2-\mu_2E_2-\delta E_2\\[5pt]
    &\frac{dS_1(t)}{dt}=\tau_1 E_1-\eta_1 S_1-\overline{k}\alpha_1S_1R_1-\overline{k}\gamma S_1S_2-\delta S_1\\[5pt]
    &\frac{dS_2(t)}{dt}=\tau_2 E_2-\eta_2 S_2-\overline{k}\alpha_2S_2R_2-\delta S_2\\[5pt]
\end{split}
\end{equation}
From the subsystem (\ref{eqn3}), we find the following transmission matrix $F$, associated with new infection term,s and transition matrix $V$, considering transferred terms, at rumor-free equilibrium  $\overline{E}_0$.

\begin{equation*}
 F=   
 \begin{bmatrix}
    0 & 0 & \overline{k}\lambda_1 & 0 \\[7pt]
    0 & 0 & 0 & \overline{k}\lambda_2 \\[7pt]
     0 & 0 & 0 & 0 \\[7pt]
      0 & 0 & 0 & 0 \\
\end{bmatrix}
\end{equation*}
and
\begin{equation*}
 V=   
 \begin{bmatrix}
    \tau_1+\mu_1+\delta & 0 & 0 & 0\\[7pt]
    0 & \tau_2+\mu_2+\delta & 0 & 0 \\[7pt]
    -\tau_1 & 0 & \eta_1+\delta & 0 \\[7pt]
    0 & -\tau_2 & 0 & \eta_2+\delta\\
\end{bmatrix}
\end{equation*}
Then
\begin{equation*}
 FV^{-1}=   
 \begin{bmatrix}
    \frac{\overline{k}\lambda_1\tau_1}{(\delta+\eta_1)(\delta+\mu_1+\tau_1)} & 0 & \frac{\overline{k}\lambda_1}{\delta+\eta_1} & 0\\[7pt]
    0 & \frac{\overline{k}\lambda_2\tau_2}{(\delta+\eta_2)(\delta+\mu_2+\tau_2)} & 0 & \frac{\overline{k}\lambda_2}{\delta+\eta_2} \\[7pt]
    0 & 0 & 0 & 0 \\[7pt]
    0 & 0 & 0 & 0\\
\end{bmatrix}
\end{equation*}
The basic reproduction number is the spectral radius of the matrix $FV^{-1}$, that is, $\mathcal{R}_0=\rho(FV^{-1})$.
\noindent Therefore, it follows that the basic reproduction number of the model is

\begin{equation*}
   \mathcal{R}_0=\text{max}\left\{\mathcal{R}^1_0,\mathcal{R}^2_0\right\}
\end{equation*}
where, $\mathcal{R}^1_0=\frac{\overline{k}\lambda_1\tau_1}{(\delta+\eta_1)(\delta+\mu_1+\tau_1)}$ and $\mathcal{R}^2_0=\frac{\overline{k}\lambda_2\tau_2}{(\delta+\eta_2)(\delta+\mu_2+\tau_2)}$

\subsection{Local stability of rumor-free equilibrium}
We establish the local stability of the rumor-free equilibrium$(\overline{E}_0)$ of the model using the Jacobian matrix of the model (\ref{eqn1}) at $\overline{E}_0$, which is given by the following.

\begin{equation*}
 J(\overline{E}_0)=   
 \begin{bmatrix}
   -\delta & 0 & 0 & \eta_1-\overline{k}\lambda_1-\overline{k}\beta_1 & \eta_2-\overline{k}\lambda_2-\overline{k}\beta_2 & 0 & 0 \\[8pt]
   
   0 & -\tau_1-\mu_1-\delta & 0 & \overline{k}\lambda_1 & 0 & 0 & 0\\[8pt]
   
   0 & 0 & -\tau_2-\mu_2-\delta & 0 & \overline{k}\lambda_2 & 0 & 0\\[8pt]
   
   0 & \tau_1 & 0 & -\eta_1-\delta & 0 & 0 & 0\\[8pt]
   
   0 & 0 & \tau_2 & 0 & -\eta_2-\delta & 0 & 0 \\[8pt]
   0 & \mu_1 & 0 & \overline{k}\beta_1 & 0 & -\delta & 0 \\[8pt]
   0 & 0 & \mu_2 & 0 & \overline{k}\beta_2 & 0 & -\delta \\
   
\end{bmatrix}
\end{equation*}
Expanding the determinant of the characteristic equation $|J-\lambda I| = 0$ by the first column, then the second last column, and then by the last column, we obtain three of the eigenvalues of $J$: $-\delta$, $-\delta$ and $-\delta$. The rest of the eigenvalues can be obtained from the eigenvalues of the $4\times 4$ matrix.
\begin{equation*}
 J_1(\overline{E}_0)=   
 \begin{bmatrix}
   -\tau_1-\mu_1-\delta & 0 & \overline{k}\lambda_1 & 0 \\[8pt]
   
   0 & -\tau_2-\mu_2-\delta & 0 & \overline{k}\lambda_2 \\[8pt]
   
   \tau_1 & 0 & -\eta_1-\delta & 0 \\[8pt]
   
   0 & \tau_2 & 0 & -\eta_2-\delta \\[8pt]
\end{bmatrix}
\end{equation*}
Then, we obtain the following characteristic polynomial of the matrix $J_1$ 
\begin{dmath}\label{eqn4}
    (\delta^2+\lambda^2+\lambda \mu_1+\lambda \tau_1-\overline{k}\lambda_1\tau_1+\eta_1\lambda+\eta_1\mu_1+\eta_1\tau_1+\eta_1\delta+2\delta\lambda+\delta\mu_1+\delta\tau_1)(-\overline{k}\lambda_2\tau_2+(\delta+\eta_2+\lambda)(\delta+\mu_2+\lambda+\tau_2))=0
\end{dmath}
So, we get the eigenvalues as
$$\lambda_{4,5}=\frac{-(\mu_1+\tau_1+\eta_1+2\delta)\pm\sqrt{(\mu_1+\tau_1+\eta_1+2\delta)^2-4(\delta^2+\eta_1\mu_1+\eta_1\tau_1+\eta_1\delta+\delta\mu_1+\delta\tau_1-\overline{k}\lambda_1\tau_1)}}{2}$$
and
$$\lambda_{6,7}=\frac{-(\mu_2+\tau_2+\eta_2+2\delta)\pm\sqrt{(\mu_2+\tau_2+\eta_2+2\delta)^2-4(\delta^2+\eta_2\mu_2+\eta_2\tau_2+\eta_2\delta+\delta\mu_2+\delta\tau_2-\overline{k}\lambda_2\tau_2)}}{2}$$

Now, for the rumor-free equilibrium to be locally asymptotically stable, we need $\lambda_{4,5}<0$ and $\lambda_{6,7}<0$. Then
    $$\lambda_{4,5}<0$$
    $$\Rightarrow  \delta^2+\eta_1\mu_1+\eta_1\tau_1+\delta\eta_1+\delta\mu_1+\delta\tau_1>\overline{k}\lambda_1\tau_1$$
    $$\Rightarrow \frac{\overline{k}\lambda_1\tau_1}{(\delta+\eta_1)(\delta+\mu_1+\tau_1)}<1$$
    $$\therefore \mathcal{R}^1_0<1$$
and
    $$\lambda_{6,7}<0$$
    $$\Rightarrow  \delta^2+\eta_2\mu_2+\eta_2\tau_2+\delta\eta_2+\delta\mu_2+\delta\tau_2>\overline{k}\lambda_2\tau_2$$
    $$\Rightarrow \frac{\overline{k}\lambda_2\tau_2}{(\delta+\eta_2)(\delta+\mu_2+\tau_2)}<1$$
    $$\therefore \mathcal{R}^2_0<1$$

As we have $\mathcal{R}^1_0<1$ and $\mathcal{R}^2_0<1$ then it implies that $\mathcal{R}_0 < 1$.

So, we have if $\mathcal{R}_0<1$, then the \emph{RFE} is locally asymptotically stable. 

\subsection{Global stability of rumor-free equilibrium}

\textbf{Theorem 3.4.1:} The rumor free equilibrium, $(\overline{E}_0)$, of the model (\ref{eqn1}) is globally asymptotically stable if $\mathcal{R}_0 <1$.\\ \\
\emph{\textbf{Proof:}} We prove the theorem by using the Lyapunov function and the LaSalle Invariance Principle \cite{hale1969ordinary,la1976stability}. Consider the following Lyapunov function
$$\mathcal{F}=f_1E_1+f_2E_2+f_3S_1+f_4S_2,$$
where
\begin{align*}
f_1&=\frac{\omega\psi\tau_1\tau_2}{(\tau_1+\mu_1+\delta)(\tau_2+\mu_2+\delta)}\\[5pt]
f_2&=\frac{\alpha_1\tau_1\tau_2}{(\tau_1+\mu_1+\delta)(\tau_2+\mu_2+\delta)}\\[5pt]
f_3&=\frac{\omega\psi\tau_2}{\tau_2+\mu_2+\delta}\\[5pt]
f_4&=\frac{\alpha_1\tau_1}{\tau_1+\mu_1+\delta}\\
\end{align*}
Here, $\omega$ is a large quantity that satisfies $\omega\psi>\alpha_1$.
Now, the time derivative of the Lyapunov function is given by
\begin{align*}
   \dot{\mathcal{F}}&=f_1\dot{E_1}+f_2\dot{E_2}+f_3\dot{S_1}+f_4\dot{S_2}\\[7pt]
   
   &=\frac{\omega\psi\tau_1\tau_2}{(\tau_1+\mu_1+\delta)(\tau_2+\mu_2+\delta)}\left(\overline{k}\lambda_1IS_1-\tau_1E_1-\overline{k}\sigma E_1S_2-\mu_1E_1-\delta E_1\right)\\[7pt]
   &\hspace{.5cm} + \frac{\alpha_1\tau_1\tau_2}{(\tau_1+\mu_1+\delta)(\tau_2+\mu_2+\delta)} \left( \overline{k}\lambda_2IS_2+\overline{k}\sigma E_1S_2+\overline{k}\gamma S_1S_2+\overline{k}\psi S_2R_1-\tau_2E_2-\mu_2E_2-\delta E_2\right)\\[7pt]
   &\hspace{1cm} + \frac{\omega\psi\tau_2}{\tau_2+\mu_2+\delta} \left( \tau_1 E_1-\eta_1 S_1-\overline{k}\alpha_1S_1R_1-\overline{k}\gamma S_1S_2-\delta S_1 \right)\\[7pt]
    &\hspace{1.5cm} + \frac{\alpha_1\tau_1}{\tau_1+\mu_1+\delta} \left(\tau_2 E_2-\eta_2 S_2-\overline{k}\alpha_2S_2R_2-\delta S_2 \right)\\[7pt]
    
   &=\frac{\omega\psi S_1\tau_2}{\tau_2+\mu_2+\delta}\left[\frac{\overline{k}\tau_1\lambda_1}{\tau_1+\mu_1+\delta}I-(\eta_1+\delta)-\overline{k}\alpha_1R_1 \right]\\[7pt]
   &\hspace{.5cm}+\frac{\alpha_1\tau_1 S_2}{\tau_1+\mu_1+\delta}\left[\frac{\overline{k}\tau_2\lambda_2}{\tau_2+\mu_2+\delta}I-(\eta_2+\delta)-\overline{k}\alpha_2R_2+\frac{\tau_2}{\tau_2+\mu_2+\delta}\overline{k}\psi R_1 \right]\\[7pt]
   &\hspace{1cm}-\frac{\tau_1\tau_2\sigma E_1S_2\overline{k}}{(\tau_1+\mu_1+\delta)(\tau_2+\mu_2+\delta)}(\omega\psi-\alpha_1)-\frac{\tau_2\gamma S_1S_2\overline{k}}{\tau_2+\mu_2+\delta}\left[ \frac{\omega\psi(\tau_1+\mu_1+\delta)-\alpha_1\tau_1}{\tau_1+\mu_1+\delta} \right]\\[7pt]
   
   &\le \frac{\omega\psi\tau_2}{\tau_2+\mu_2+\delta}\left[\frac{\overline{k}\tau_1\lambda_1}{\tau_1+\mu_1+\delta}I-(\eta_1+\delta)-\overline{k}\alpha_1R_1 \right]\\[7pt]
   &\hspace{.5cm}+\frac{\alpha_1\tau_1}{\tau_1+\mu_1+\delta}\left[\frac{\overline{k}\tau_2\lambda_2}{\tau_2+\mu_2+\delta}I-(\eta_2+\delta)-\overline{k}\alpha_2R_2+\frac{\tau_2}{\tau_2+\mu_2+\delta}\overline{k}\psi R_1 \right]\\[7pt]
   &\hspace{1cm}-\frac{\tau_1\tau_2\sigma E_1S_2\overline{k}}{(\tau_1+\mu_1+\delta)(\tau_2+\mu_2+\delta)}(\omega\psi-\alpha_1)-\frac{\tau_2\gamma S_1S_2\overline{k}}{\tau_2+\mu_2+\delta}\left[ \frac{(\omega\psi-\alpha_1)\tau_1+\omega\psi(\mu_1+\delta)}{\tau_1+\mu_1+\delta} \right]\\[7pt]
   
   & \le \frac{\omega\psi\tau_2(\eta_1+\delta)}{\tau_2+\mu_2+\delta} \left [ \frac{\tau_1\lambda_1\overline{k}}{(\eta_1+\delta)(\tau_1+\mu_1+\delta)}-1 \right]+\frac{\alpha_1\tau_1(\eta_2+\delta)}{\tau_1+\mu_1+\delta} \left [ \frac{\tau_2\lambda_2\overline{k}}{(\eta_2+\delta)(\tau_2+\mu_2+\delta)}-1\right]\\[7pt]
   &\hspace{.5cm} -\frac{\alpha_1\tau_1}{\tau_1+\mu_1+\delta}\alpha_2R_2\overline{k}-\frac{\overline{k}\psi\tau_2\alpha_1R_1}{\tau_2+\mu_2+\delta} \left [ \omega-\frac{\tau_1}{\tau_1+\mu_1+\delta} \right]-\frac{\overline{k}\tau_1\tau_2\sigma E_1S_2}{(\tau_1+\mu_1+\delta)(\tau_2+\mu_2+\delta)}(\omega\psi-\alpha_1)\\[7pt]
   & \hspace{1cm} -\frac{\overline{k}\tau_2\gamma S_1S_2}{\tau_2+\mu_2+\delta}\frac{(\omega\psi-\alpha_1)\tau_1+\omega\psi(\mu_1+\delta)}{\tau_1+\mu_1+\delta}\\[7pt]
   
   & \le \frac{\omega\psi\tau_2(\eta_1+\delta)}{\tau_2+\mu_2+\delta} \left [ \mathcal{R}^1_0-1 \right]+\frac{\alpha_1\tau_1(\eta_2+\delta)}{\tau_1+\mu_1+\delta} \left [ \mathcal{R}^2_0-1\right]-\frac{\alpha_1\tau_1\alpha_2R_2\overline{k}}{\tau_1+\mu_1+\delta}-\frac{\overline{k}\psi\tau_2\alpha_1R_1}{\tau_2+\mu_2+\delta} \left [ \frac{(\omega-1)\tau_1+\omega(\mu_1+\delta)}{\tau_1+\mu_1+\delta} \right]\\[7pt]
   &\hspace{1cm}-\frac{\overline{k}\tau_1\tau_2\sigma E_1S_2}{(\tau_1+\mu_1+\delta)(\tau_2+\mu_2+\delta)}(\omega\psi-\alpha_1)-\frac{\overline{k}\tau_2\gamma S_1S_2}{\tau_2+\mu_2+\delta}\frac{(\omega\psi-\alpha_1)\tau_1+\omega\psi(\mu_1+\delta)}{\tau_1+\mu_1+\delta}\\
\end{align*}
This gives, $\mathcal{\dot{F}}<0$ if $\mathcal{R}^1_0 < 1$ and $\mathcal{R}^2_0 < 1$. According to LaSalle’s invariance principle, therefore, it implies that $\overline{E}_0$ is globally asymptotically stable for $\mathcal{R}_0 < 1$.

\subsection{Global stability of rumor prevailing equilibrium}
From the expression that we have calculated for $\mathcal{R}^1_0$ and $\mathcal{R}^2_0$ with the help of the results obtained in [81, 82],  we can derive that no matter what value of $\alpha_{1},\alpha_{2},\beta_{1},\beta_{2},\sigma,\gamma,  $ and $\psi$ takes, the threshold value of rumor will
not change. Therefore, to simplify the calculation, we consider $\alpha_{1}=0,\alpha_{2}=0,\beta_{1}=0,\beta_{2}=0,\sigma=0,\gamma=0,  $ and $\psi=0$  in the following analysis. Then, we obtain the following system.

\begin{equation}\label{eqnp1}
\begin{split}
    & \xi=-\eta_1S_1^{*}-\eta_2S_2^{*}+\overline{k}\lambda_1I^{*}S_1^{*}+\overline{k}\lambda_2I^{*}S_2^{*}+\delta I^{*}\\[5pt]
    & \overline{k}\lambda_1I^{*}S_1^{*}-\tau_1E_1^{*}-\mu_1E_1^{*}-\delta E_1^{*}=0\\[5pt]
    & \overline{k}\lambda_2I^{*}S_2^{*}-\tau_2E_2^{*}-\mu_2E_2^{*}-\delta E_2^{*}=0\\[5pt]
    &\tau_1 E_1^{*}-\eta_1 S_1^{*}-\delta S_1^{*}=0\\[5pt]
    &\tau_2 E_2^{*}-\eta_2 S_2^{*}-\delta S_2^{*}=0\\[5pt]
    &\mu_1E_1^{*}-\delta R_1^{*}=0\\[5pt]
    &\mu_2E_2^{*}-\delta R_2=0.\\[5pt]
\end{split}
\end{equation}

We denote that the rumor prevailing equilibrium points of the model as
$\overline{E}^{*}=(I^*, E_1^{*}, E_2^{*},S_1^{*},S_2^{*},R_1^{*}, R_2^{*}).$

Our investigation of the global stability of the equilibrium state $\overline{E}^*$ is constrained under a specific condition where $\frac{I}{I^*} = \frac{S_1}{S_1^*}$. Consequently, we put forth the following statement.

\textbf{Theorem 3.4.2:} The unique rumor prevailing equilibrium point, $\overline{E} ^{*}$, of the model \eqref{eqn1} is globally asymptotically stable if $\mathcal{R}_0>1$ and $\frac{I}{I ^{*}}$ = $\frac{S_1}{S_1 ^{*}}$.\\ \\
\emph{\textbf{Proof:}} Consider the following Lyapunov function. 
\begin{equation} \label{lya}
     \mathcal{F}=G_1 I^{*} g(\frac{I(t)}{I^{*}})+G_2 E_1^{*}g(\frac{E_1(t)}{E_1^{*}})+G_3E_2^{*}g(\frac{E_2(t)}{E_2 ^{*}})+G_4 S_1^{*} g(\frac{S_2(t)}{S_2^{*}})+G_5 S_2^{*}g(\frac{S_2(t)}{S_2 ^{*}}) 
 \end{equation}
and $g(x)=x-1-ln(x), \text{ for all } x>0$. We choose $G_{1}=1,G_{2}=1,G_{3}=1,G_{4}=1, G_{5}=1.$ Now, set 
 $$ x_1(t)=\frac{I(t)}{I^{*}}, x_2(t)=\frac{E_{1}(t)}{E_1^{*}}, x_3(t)=\frac{E_{2}(t)}{E_2^{*}} ,x_4(t)=\frac{S_{1}(t)}{S_1^{*}},x_5(t)=\frac{S_{2}(t)}{S_2^{*}}
$$
For convenient, in the following analysis, we denote $x_1(t)=x_1,x_2(t)=x_2,x_3(t)=x_3,x_4(t)=x_4,x_5(t)=x_5$. The model \eqref{eqn1} and the system \eqref{eqnp1} yields the following.

\begin{equation} \label{pp1}
\begin{split}
& \frac{dI(t)}{dt} =  k\lambda_1 I^{*}S_1^{*}(1-x_1x_4)+k\lambda_2 I S_2^{*}(1-x_1x_5)
+\delta I^{*}(1-x_1)-\eta_1 S_1 ^{*}(1-x_4)-\eta_2 S_2 ^{*}(1-x_5)\\[5pt]
& \frac{dE_1(t)}{dt} =  -k\lambda_1 S_1^{*}(1-x_4)+\tau_1 E_1^{*} (1-x_2) +\mu_1 E_1^{*} (1-x_2) +\delta E_1^{*} (1-x_2)\\[5pt]
& \frac{dE_2(t)}{dt}  = -k \lambda_2 I^{*}S_2{*}(1-x_1 x_5)+\tau_2E_2^{*}(1-x_3)+\mu_2 E_2^{*}(1-x_3)+\delta E_2^{*}(1-x_3)\\[5pt]
&\frac{dS_1(t)}{dt} =  -\tau_1 E_1^{*}(1-x_2)+\eta_1 S_1^{*} (1-x_4)+\delta S_1^{*}(1-x_4) \\[5pt]
&\frac{dS_2(t)}{dt} =  \tau_2 E_2^{*}(1-x_3)+\eta_2 S_2^{*} (1-x_5)+\delta S_2^{*}(1-x_5) \\[5pt]
&\frac{dR_1(t)}{dt} = -\mu_1 E_1^{*}(1-x_2)+\delta R_1^{*}(1-x_6) \\[5pt]
&\frac{dR_2(t)}{dt}  = \mu_2 E_2^{*}(1-x_3)+\delta R_2^{*}(1-x_7) \\[5pt]
\end{split}
\end{equation}
From \eqref{pp1} using the definition of $g(x)$, the derivative of $\mathcal{F}(t)$ can be derived as.
\begin{align*}
\dot{\mathcal{F}}&=G_1 \dot{I}+G_2\dot{E_1}+G_2\dot{E_2}+G_3\dot{S_1}+G_4\dot{S_2}\\[7pt]
    &= (1-\frac{1}{x_1})[k\lambda_1 I^{*}S_1^{*}(1-x_1x_4)+k\lambda_2 I S_2^{*}(1-x_1x_5)
+\delta I^{*}(1-x_1)-\eta_1 S_1 ^{*}(1-x_4)-\eta_2 S_2 ^{*}(1-x_5)]\\
&+(1-\frac{1}{x_2})[ -k\lambda_1 S_1^{*}(1-x_4)+\tau_1 E_1^{*} (1-x_2) +\mu_1 E_1^{*} (1-x_2) +\delta E_1^{*} (1-x_2)] \\
&+(1-\frac{1}{x_3}) [-k \lambda_2 I^{*}S_2{*}(1-x_1 x_5)+\tau_2E_2^{*}(1-x_3)+\mu_2 E_2^{*}(1-x_3)+\delta E_2^{*}(1-x_3)]\\
&+(1-\frac{1}{x_4})[-\tau_1 E_1^{*}(1-x_2)+\eta_1 S_1^{*} (1-x_4)+\delta S_1^{*}(1-x_4)]\\
&+(1-\frac{1}{x_5})\tau_2 E_2^{*}(1-x_3)+\eta_2 S_2^{*} (1-x_5)+\delta S_2^{*}(1-x_5) 

\end{align*}
Simplifying the above equation and using system \eqref{eqnp1}, we further find the following.
\begin{align*}
    \frac{d\mathcal{F}(t)}{dt}&=g(x_4)[k\lambda_1 I^{*} S_1^{*}-\delta S_1^{*}]-\xi g(x_1)+g(x_1)[-\delta I^{*}+k\lambda_1 S_1^{*}]-\eta_1 S_1^{*}g(\frac{x_4}{x_1})-\eta_2 S_2^{*}g(\frac{x_5}{x_1})\\
    &-2k \lambda_1 S_1^{*}g(\frac{x_1}{x_2})-k \sigma E_1^{*}g(x_2 x_4)-\tau_1 E_1 ^{*} g(\frac{x_2}{x_1})
\end{align*}
Using the condition  $\frac{I}{I ^{*}}$ = $\frac{S_1}{S_1 ^{*}}$ we have, 
\begin{equation} \label{pp8}
     \frac{d\mathcal{F}(t)}{dt} =-\xi g(x_1)-\eta_1 S_1^{*}g(\frac{x_4}{x_1})-\eta_2 S_2^{*}g(\frac{x_5}{x_1})\\
    -2k \lambda_1 S_1^{*}g(\frac{x_1}{x_2})-k \sigma E_1^{*}g(x_2 x_4)-\tau_1 E_1 ^{*} g(\frac{x_2}{x_1})
\end{equation}

Obviously, Eq.~\eqref{pp8} implies that $\frac{d \mathcal{F}}{dt}< 0$ whenever  $\frac{I}{I ^{*}}$ = $\frac{S_1}{S_1 ^{*}}$ and $\mathcal{R}_0>1$.

Therefore, the rumor prevailing equilibrium of the model \eqref{eqn1} is globally asymptotically stable if $\mathcal{R}_{0}>1$ and $\frac{I}{I ^{*}}$ = $\frac{S_1}{S_1 ^{*}}$.
\section{Optimal control of the model}
We use the optimal control strategy to control the spread of rumors in the network.
Our objective is to reduce the spread of rumors by moving individuals in $E_1, E_2, S_1$, and $S_2$ classes to $R_1$ and $R_2$ classes.This is because people exposed to rumors may later become spreaders, further spreading the rumors throughout the network. Since it's difficult to prevent people from hearing information, we assume that everyone in the network will eventually be exposed to rumors. Therefore, our focus is on influencing those who have already been exposed and preventing them from becoming spreaders themselves. At the same time, we need to identify and stop current spreaders from spreading the rumors. To achieve this, we aim to identify the spreaders, impose fines, and incarcerate them as appropriate.
To achieve this goal, we have four different strategies. The first strategy, $u_1(t)$, involves the authority releasing an official statement about the true nature of the first rumor being spread. This statement will be circulated through various mediums, such as television, radio, and online news portals. The second strategy, $u_2(t)$, has a two-fold approach. Firstly, it helps identify the true nature of the second piece of information. Secondly, if this information is deemed to be a rumor, the strategy works towards influencing individuals in the social network not to believe the rumor and prevent its spread. To implement this, the authority will employ trained professionals to establish algorithms to identify the true nature of the second information being disseminated. Suppose the information is proven to be a rumor. In that case, they will hire famous "social media" influencers to create awareness-raising online content and help make this content go viral in online social media. The authority will ensure that these contents contain enough clarification about the true nature of the rumor. The third strategy, $u_3(t)$, is where the authorities intend to penalize individuals who actively spread the first rumor. By taking away a heavy amount of money from the spreaders for spreading wrong information, in the future, they will be cautious and stop spreading rumors and move into the $R_1$ class. The fourth strategy, $u_4(t)$, is a more complex strategy, similar to $u_2(t)$. In this case, the authorities first identify the spreaders of the second rumor by seeking help from experts. After determining the spreaders, the authorities take steps to incarcerate them for a specific amount of time and share this event of incarceration throughout the social network.
The system of differential equations governing the controlled mathematical model for \eqref{eqn1} is expressed as follows.
\begin{equation}\label{eqnp2}
\begin{split}
    & \frac{dI(t)}{dt}=\xi+\eta_1S_1+\eta_2S_2-\overline{k}\lambda_1IS_1-\overline{k}\lambda_2IS_2-\overline{k}\beta_1IS_1-\overline{k}\beta_2IS_2-\delta I\\[5pt]
    & \frac{dE_1(t)}{dt}=\overline{k}\lambda_1IS_1-\tau_1E_1-\overline{k}\sigma E_1S_2-\mu_1E_1-\delta E_1-u_{1}E_{1}\\[5pt]
    & \frac{dE_2(t)}{dt}=\overline{k}\lambda_2IS_2+\overline{k}\sigma E_1S_2+\overline{k}\gamma S_1S_2+\overline{k}\psi S_2R_1-\tau_2E_2-\mu_2E_2-\delta E_2-u_{2}E_{2}\\[5pt]
    &\frac{dS_1(t)}{dt}=\tau_1 E_1-\eta_1 S_1-\overline{k}\alpha_1S_1R_1-\overline{k}\gamma S_1S_2-\delta S_1-u_{3}S_{1}\\[5pt]
    &\frac{dS_2(t)}{dt}=\tau_2 E_2-\eta_2 S_2-\overline{k}\alpha_2S_2R_2-\delta S_2-u_{4}S_{2}\\[5pt]
    &\frac{dR_1(t)}{dt}=\mu_1E_1+\overline{k}\beta_1IS_1+\overline{k}\alpha_1S_1R_1-\overline{k}\psi S_2R_1-\delta R_1+u_{1}E_{1}+u_{3}S_{1}\\[5pt]
    &\frac{dR_2(t)}{dt}=\mu_2E_2+\overline{k}\beta_2IS_2+\overline{k}\alpha_2S_2R_2-\delta R_2+u_{2}E_{2}+u_{4}S_{2}\\[5pt]
\end{split}
\end{equation}
With initial conditions.
\begin{equation} \label{initial for oc}
I(0)=I_0 \geq 0 , E_{1}(0)=E_{10} \geq 0 , E_2(0)=E_{20} \geq 0 , S_1(0)=S_{10} \geq 0  , S_2(0)=S_{20} \geq 0,
\end{equation}
$$R_1(0)=R_{10} \geq 0  , R_2(0)=R_{20} \geq 0$$
Our primary goal is to minimize the maximum number of people who believe the rumors and disseminate them in the network afterward. At the same time, we want to reduce the cost of applying our optimal control strategies given the initial population sizes of all seven classes $I_0, E_{10}, E_{20}, S_{10}, S_{20}, R_{10}$ and $R_{20}$. The objective function of our optimal control problem is established as follows.
\begin{align}
\label{eqn:object}
\nonumber
\mathcal{J}(u_{1} (t),u_{2} (t),u_{3}(t),u_{4}(t)) =&  \int_0^T \Biggl [ A_{1} I_1(t)+A_{2} E_1(t)+A_{3} S_1(t)+A_{4}S_2(t)\\
& + \frac{1}{2} \Biggl \{ B_{1}(u_{1})^2 +B_{2}(u_{2})^2 +B_{3}(u_{3})^2+B_{4}(u_{4})^2\Biggl \}  \Biggl]dt,
\end{align}
subject to the system of equation \eqref{eqnp2} while control set $U$ is Lebesgue measurable, which is defined as follows.

\begin{align}
U= \left\{ u_{i}(t)\mid   u_{i} \text{ is piecewise continuous function, } 0 \leq u_{i}(t) \leq 1 \text{  and }t\in [0,t_f] \right\}.
\end{align}

In equation \eqref{eqn:object}, $A_{1}, A_{2}, A_{3},\text{ and }A_{4}$ are positive balancing coefficients of the exposed, and infected individual's densities $E_1(t), E_2(t), S_1(t)$ and $S_2 (t)$ respectively. $B_{1}, B_{2}$, $B_{3}$ and $B_4$ are the positive weight coefficients of control costs. 

\subsection{Existence of an optimal control }
To prove Existence, we will use prove Theorem 4.1 of chapter three from Fleming and Rishel \cite{fleming2012deterministic}. The proofs are given below: 

\textbf{Proof: } (i)\\
   To prove that $\mathcal{F}$ is nonempty, let
\begin{align*}
&\frac{dI}{dt}=F_{1}(t,E_1,E_2,S_1,S_2,R_1,R_2)\\
&\frac{dE_1}{dt}=F_{2}(t,E_1,E_2,S_1,S_2,R_1,R_2)\\
&\frac{dE_2}{dt}=F_{3}(t,E_1,E_2,S_1,S_2,R_1,R_2)\\
&\frac{dS_1}{dt}=F_{4}(t,E_1,E_2,S_1,S_2,R_1,R_2)\\
&\frac{dS}{dt}=F_{5}(t,E_1,E_2,S_1,S_2,R_1,R_2)\\
&\frac{dR_1}{dt}=F_{6}(t,E_1,E_2,S_1,S_2,R_1,R_2)\\
&\frac{dR_2}{dt}=F_{7}(t,E_1,E_2,S_1,S_2,R_1,R_2)\\
\end{align*}
where $ F_{1},F_{2},F_{3},F_{4},F_{5 },F_6,F_7 $ form the RHS of system\eqref{eqnp2}. Let, $u_{1}(t)=C_{1},u_{2}(t)=C_{2},u_{3}(t)=C_{3}, u_4=C_4$ for some constant $C_1,C_2,C_3,C_4$. $F_{1}, F_{2}, F_{3}, F_{4},F_5,F_6 $ and $F_{7}$ are linear. Thus, they are continuous everywhere. Additionally, the partial derivatives of $F_{1}, F_{2}, F_{3}, F_{4}.F_5, F_6$, and $F_{7}$ with respect to all states are constants, and so they are also continuous
everywhere. Therefore, there exists a unique solution ${I(t)=\sigma_{1},E_1(t)=\sigma_{2},E_2(t)=\sigma_{3},S_1(t)=\sigma_{4},S_2(t)=\sigma_{5}, R_1(t)=\sigma_6, R_2(t)=\sigma_7}$ that satisfies the initial conditions. Therefore, the set of controls and corresponding state variables is nonempty, and condition (i) is satisfied\\ 

{\bf Proof (iii):}\\
     By definition, U is closed. Take any controls $ u_{1},u_{2}  \in U $ and $\epsilon \in [0,1]$ such that 
    $$ 0 \leq \epsilon u_{1}+(1-\epsilon)u_{2}$$

$$
\epsilon u_1+(1-\epsilon)u_2 \leq \epsilon+(1-\epsilon)=1
$$ 
Hence $0 \leq \epsilon u_{1}+(1-\epsilon)u_{2} \leq 1$ for all $u_{1},u_{2} \in U \hspace{.2cm} and\hspace{.2cm} \epsilon \in [0,1]$.\\
So U is convex, and therefore, condition (ii) is satisfied.\\

{\bf Proof (iii):} If we consider for any k, 
$$ 
F_{1} \leq  \xi+\eta_1S_1+\eta_2S_2\\
$$
$$
F_{2} \leq  \overline{k}\lambda_1 IS_1-u_{1}E_{1}\\
$$
$$
F_{3} \leq  \overline{k}\lambda_2IS_2+\overline{k}\sigma E_1S_2+\overline{k}\gamma S_1S_2+\overline{k}\psi S_2R_1-u_{2}E_{2}\\
$$
$$
F_{4} \leq  \tau_1 E_11S_2-u_{3}S_{1}\\
$$
$$
F_{5} \leq   \tau_2 E_2-u_{4}S_{2}\\
$$
$$
F_{6} \leq   \mu_1E_1+\overline{k}\beta_1IS_1+\overline{k}\alpha_1S_1R_1+u_{3}S_{1}\\
$$
$$
F_{6} \leq   \mu_2E_2+\overline{k}\beta_2IS_2+\overline{k}\alpha_2S_2R_2+u_{4}S_{2}\\
$$

Then, the following system 
$$ 
\frac{dI}{dt}=F_{1}(t,I,E_1,E_2,S_1,S_2,R_1,R_2)\\
$$
$$
\frac{dE_1}{dt}=F_{2}(t,I,E_1,E_2,S_1,S_2,R_1,R_2)\\
$$
$$
\frac{dE_2}{dt}=F_{3}(t,I,E_1,E_2,S_1,S_2,R_1,R_2)\\
$$
$$
\frac{dS_1}{dt}=F_{4}(t,I,E_1,E_2,S_1,S_2,R_1,R_2)\\
$$
$$
\frac{dE_2}{dt}=F_{5}(t,I,E_1,E_2,S_1,S_2,R_1,R_2)\\
$$ 
$$
\frac{dR_1}{dt}=F_{6}(t,I,E_1,E_2,S_1,S_2,R_1,R_2)\\
$$ 
$$
\frac{dR_2}{dt}=F_{7}(t,I,E_1,E_2,S_1,S_2,R_1,R_2)\\
$$ 

can be written as 
\begin{center}
\begin{equation} \label{matr}
\begin{split}
\overline{F}(t,\overline{X},u) \leq \overline{m} \begin{pmatrix}
t, &\begin{bmatrix}
I(t)\\
E_{1}(t) \\
E_{2}(t)\\
S_{1}(t) \\
S_{2}(t) \\
R_{1}(t) \\
R_{2}(t) 
\end{bmatrix} 
\end{pmatrix} \overline{X}(t)+\overline{n}\begin{pmatrix}
t, & \begin{bmatrix}
I(t)\\
E_{1}(t) \\
E_{2}(t)\\
S_{1}(t) \\
S_{2}(t) \\
R_{1}(t) \\
R_{2}(t) 
\end{bmatrix} 
\end{pmatrix} \begin{pmatrix}
u_{1}(t)\\
u_{2}(t)\\
u_{3}(t)\\
u_{4}(t)
\end{pmatrix}
\end{split}
\end{equation}
\end{center}

where,

\begin{center}
\begin{equation} \label{eqx}
\begin{split}
\overline{m} \begin{pmatrix}
t,& \begin{bmatrix}
I(t)\\
E_{1}(t) \\
E_{2}(t)\\
S_{1}(t) \\
S_{2}(t) \\
R_{1}(t) \\
R_{2}(t) 
\end{bmatrix} 
\end{pmatrix} = \begin{bmatrix}
0&0&0&\eta_1&\eta_2&0&0\\
\overline{k}\lambda_1 S_{1_{max}}&0&0&0&0&0&0\\
\overline{k}\lambda_2 S_{2_{max}}&\overline{k}\sigma S_{2_{max}}&0&\overline{k}\gamma S_{2_{max}}&0&\overline{k} \psi S_{2_{max}}&0\\
0\tau_1&0&0&0&0&0&0 \\
0&0&\tau_2&0&0&0&0\\
\overline{k}\beta_1S_{1_{max}}&\mu_1&0&0&0&\overline{k}\alpha_1S_{1_{max}}&0\\
\overline{k}\beta_2 S_{2_{max}}&0&\mu_2&0&0&\overline{k}0&\alpha_2 S_{2_{max}}
\end{bmatrix} 
\end{split}
\end{equation}
\end{center}
and, 
\begin{center}
\begin{equation} \label{qqqq}
\begin{split}
\overline{n} \begin{pmatrix}
t,& \begin{bmatrix}
I(t)\\
E_{1}(t)\\
E_{2}(t) \\
S_{1}(t)\\
S_{2}(t) \\
R_{1}(t)\\
R_2(t)
\end{bmatrix} 
\end{pmatrix} = \begin{bmatrix}
0\\
-E_1\\
-E_{2}\\
-S_{1} \\
S_2\\
E_1+S_1\\
E_2+S_2
\end{bmatrix} 
\end{split}
\end{equation}
\end{center}
which gives a linear function of $u_{1},u_{2},u_{3},u_4$ with coefficients determined by time and state variables.
We can then determine the bound of the RHS. Note that all parameters are constant and greater than or equal to zero. So we can write,
\begin{align*}
    \mid \overline{F}(t)(\overline{X},u_{1},u_{2},u_{3},u_{4}) \mid \leq & \lvert \mid \overline{m} \mid \rvert \mid \overline{X}\mid +\mid \overline{E}_{1_k}+\overline{E}_{2_k}+\overline{S}_1+\overline{S}_2  \mid\mid(u_{1},u_{2},u_{3},u_4) \mid \\
    & \leq G \bigg( \mid \overline{X} \mid + \mid (u_{1},u_{2},u_{3},u_{4}) \mid\bigg)
\end{align*}
where $ \overline{E}_1,\overline{E}_2,\overline{S}_1,\overline{S}_2$ are bounded, and G incorporates the upper bound of the constant matrix. Thus, we
see that the RHS is bounded by a sum of the state and the bounded control. Condition
(iii) is now satisfied.

{\bf Proof (iv):}\\
 Let $f(u_{1},u_{2},u_{3},u_{4})= A_{1} E_1(t)+A_{2} E_2(t)+A_{3} S_1(t)+A_4 S_2+\frac{1}{2} \Bigl \{ B_{1}(u_{1})^2 +b_{2}(u_{2})^2 +b_{3}(u_{3}+b_4 (u_4)^2)^2\Bigl \}$ be the integrand of the objective functional. Take $u_{1},u_{2},u_{3}, u_4 \in U $ and 
\\

$\frac{b_{1}}{2}(u_{1})^{2}-\frac{b_{1}}{2}2u_{1}u_{1}^{*}+\frac{b_{1}}{2}(u_{1}^{*})^{2}+\frac{b_{2}}{2}(u_{2})^{2}-\frac{b_{2}}{2}2u_{2}u_{2}^{*}+\frac{b_{2}}{2} (u_{2}^{*})^{2}+\frac{b_{3}}{2}(u_{3})^{2}-\frac{b_{3}}{2}2u_{3}u_{3}^{*}+\frac{b_{3}}{2} (u_{3}^{*})^{2}+\frac{b_{4}}{2}(u_{4})^{2}-\frac{b_{4}}{2}2u_{4}u_{4}^{*}+\frac{b_{4}}{2} (u_{4}^{*})^{2}  =  \frac{b_{1}}{2}(u_{1}-u_{1}^{*})^2+\frac{b_{2}}{2}(u_{2}-u_{2}^{*})^2 +\frac{b_{3}}{2}(u_{3}-u_{3}^{*})^2 +\frac{b_{4}}{2}(u_{4}-u_{4}^{*})^2 \geq 0 $ \\

$$ \implies \epsilon f(u_{1},u_{2},u_{3},u_{4}) +(1-\epsilon)f(u_{1}^{*},u_{2}^{*},u_{3}^{*},u_4^{*})\geq f(\epsilon u_{1}+(1-\epsilon)u_{1}^{*},\epsilon u_{2}+(1-\epsilon)u_{2}^{*},\epsilon u_{3}+(1-\epsilon)u_{3}^{*}),\epsilon u_{4}+(1-\epsilon)u_{4}^{*})$$\\
\\
Thus $f(u_{1},u_{2},u_{3},u_4)$ is convex on U, and now we have to prove that $\mathcal{J}(u_{1},u_{2},u_{3},u_4) \geq-Q_2+Q_1 \lvert (u_{1},u_{2},u_{3},u_4) \rvert^{\eta}$ with $Q_1>0, Q_2>0$ and $\eta>1$.
where $Q_2>0$ depends on the upper bound of $E_1(t),E_2(t),S_1(t),S_2(t)$. We can also see that $\eta=2>1, Q_1>0$. Hence, condition (iv) is satisfied. From the above observation, the Existence of the objective function has been established.

\subsection{Characterization of the Optimal Control} \label{Characterization of the Optimal Control}
In order to derive the necessary condition for optimal control, we use Pontryagin's Maximum principal~\cite{ morais2018modeling}.
\textbf{Theorem 3.4.4:}
    Let,$u_{11}^{*}, u_{2}^{*},u_{3}^{*},u_{4}^{*}$ be the optimal control variables of the system and $(I_1^{*},E_1^{*},E_2^{*},S_1^{*},S_2^{*},R_1^{*},R_2^{*}))$ be the corresponding state solutions of the system \eqref{eqnp1}, then there must be some adjoint variables $\lambda_{I}(t), \lambda_{E_1}(t), \lambda_{E_2}(t), \lambda_{S_1}(t),\lambda_{S_2}(t),\lambda_{R_1}(t),\lambda_{R_2}(t)$, which satisfy the following adjoin equation,

  \begin{center}
\begin{equation} \label{eq3221}
\begin{split}
&\lambda_{I}^{'}=\lambda_I(\overline{k}\lambda{1}S_1+\overline{k}\lambda_2S_2+\overline{k}\beta_1 S_1+\overline{k}\beta_2S_2+\delta)-\lambda_{E_1}\overline{k}\lambda_1 S_1-\lambda_{E_2}\overline{k}\lambda_2S_2-\lambda_{R_1}\overline{k}\beta_1 S_1-\lambda_{R_2}\overline{k}\beta_2 S_2 \\
& \lambda_{E_1}^{'}=-A_1 +\lambda_{E_1}(\tau_1++\overline{k}\sigma S_2 +\mu_1 +\delta)+E_1 u_1-\lambda_{E_2} \overline{k}\sigma S_2-\lambda_{S_1}\tau_1-\lambda_{R_1}\mu_1\\
& \lambda_{E_2}^{'}=-A_2+\lambda_{E_2}(\tau_2+\mu_2+\delta+u_2)-\tau_2\lambda_{S_2}-\lambda_{R_2}(\mu_2 +u_2)\\
& \lambda_{S_1}^{'}= -A_{3}-\lambda_{I}(\eta_1 -\overline{k}\lambda_1 I-\overline{k}\beta_1 I)-\lambda_{E_1}\overline{k}\lambda_1 I+ \lambda_{S_1}(\eta_1 +k \alpha_1 R_1 +\overline{k}\gamma S_2+\delta+u_3)-\lambda_{R_1}(\overline{k}\beta_1 I+\overline{k}\alpha_1 R_1 +u_3)\\
& \lambda_{S_2}^{'} = -A_4- \lambda_I (\eta_2 -\overline{k} \lambda_2 I -\overline{k}\beta_2 I) +\lambda_{E_1} \overline{k} \sigma E_1 -\lambda_{E_2}(\overline{k}\lambda_2 I+\overline{k}\sigma E_1 +\overline{k} \psi R_1)+\lambda_{S_1} \overline{k} \gamma S_1-\lambda_{S_2}(\eta_2+\overline{k} \alpha_2 R_2 +\delta+u_4)\\
&\hspace{3cm}-\lambda_{R_1}\overline{k} \psi R_1+ \lambda_{R_2}(\overline{k}\beta_2 I +\overline{k} \alpha_2 R_2 +u_4)\\
& \lambda_{R_1}^{'}=-\lambda _{E_2} \overline{k} \psi S_2 -\lambda _{S_1} \overline{k} \alpha_1 S_1-\lambda_{R_1}(\overline{k} \alpha_1 S_1 -\overline{k} \psi S_2 -\delta)\\
& \lambda_{R_2}^{'}=-\lambda_{R_2}(\overline{k} \alpha_2 S_2 -\delta)
\end{split}
\end{equation}
\end{center}
\begin{align}
 \lambda_{I_1}^{'}(T)= \lambda_{E_1}^{'}(T)= \lambda_{E_2}^{'}(T)= \lambda_{S_1}^{'}(T)= \lambda_{S_2}^{'}(T)=\lambda_{R_1}^{'}(T)= \lambda_{R_2}^{'}(T)=0
\end{align}
In addition, the corresponding optimal controls are given as follows, 

\begin{align} \label{hamu1}
    u_{1}^{*} = \min\bigg(1,\max(0,\frac{(\lambda_{E_1}-\lambda_{R_1})\Tilde{E}_{1}}{b_{1}})\bigg)
\end{align}
\begin{align} \label{hamu2}
    u_{2}^{*}=\min\bigg(1,\max(0,\frac{(\lambda_{E_2}-\lambda_{R_2})\Tilde{E}_{2}}{b_{2}})\bigg)
\end{align}
\begin{align} \label{hamu3}
    u_{3}^{*}=\min\bigg(1,\max(0,\frac{(\lambda_{S_1}-\lambda_{R_1})\Tilde{S}_{1}}{b_{3}})\bigg)
\end{align}
\begin{align} \label{hamu3}
    u_{4}^{*}=\min\bigg(1,\max(0,\frac{(\lambda_{S_2}-\lambda_{R_2})\Tilde{S}_{2}}{b_{4}})\bigg)
\end{align}
   
\textbf{Proof:} we now derive the Hamiltonian, which is given below,

\begin{center}
\begin{eqnarray}\label{eqnqw}  \nonumber
\begin{aligned}
&\mathcal{H}(I_1,E_1,E_2,S_1,S_2,u_{1},u_{2},u_{3},u_4)=\mathcal{L}(E_1,E_2,S_1,S_2,u_{1},u_{2},u_{3},u_4)\\
&+\lambda_{I}\Bigl \{ \xi+\eta_1S_1+\eta_2S_2-\overline{k}\lambda_1IS_1-\overline{k}\lambda_2IS_2-\overline{k}\beta_1IS_1-\overline{k}\beta_2IS_2-\delta I\Bigl \} \\
&+\lambda_{E_1}\Bigl \{ \overline{k}\lambda_1IS_1-\tau_1E_1-\overline{k}\sigma E_1S_2-\mu_1E_1-\delta E_1-u_{1}E_{1}\Bigl \} \\
& +\lambda_{E_2} \Bigl \{\overline{k}\lambda_2IS_2+\overline{k}\sigma E_1S_2+\overline{k}\gamma S_1S_2+\overline{k}\psi S_2R_1-\tau_2E_2-\mu_2E_2-\delta E_2-u_{2}E_{2} \Bigl \}\\
& +\lambda_{S_1}\Bigl \{ \overline{k}\lambda_2IS_2+\overline{k}\sigma E_1S_2+\overline{k}\gamma S_1S_2+\overline{k}\psi S_2R_1-\tau_2E_2-\mu_2E_2-\delta E_2-u_{2}E_{2} \Bigl\} \\
&+\lambda_{S_2}\Bigl \{ \tau_2 E_2-\eta_2 S_2-\overline{k}\alpha_2S_2R_2-\delta S_2-u_{4}S_{2}\Bigl\} \\
& +\lambda_{R_1}\Bigl \{ \mu_1E_1+\overline{k}\beta_1IS_1+\overline{k}\alpha_1S_1R_1-\overline{k}\psi S_2R_1-\delta R_1+u_{1}E_{1}+u_{3}S_{1}\Bigl\} \\
&+\lambda_{R_2}\Bigl \{\mu_2E_2+\overline{k}\beta_2IS_2+\overline{k}\alpha_2S_2R_2-\delta R_2+u_{2}E_{2}+u_{4}S_{2}\Bigl\} 
\end{aligned}
\end{eqnarray}
\end{center}

 Here, $\lambda_{E_1},\lambda_{E_2},\lambda_{S_1},\lambda_{S_2}$ denote the corresponding adjoints for the state variables $E_1,E_2,S_1,S_2$, respectively. To obtain the differential equation for the associated adjoint, we differentiate the Hamiltonian ($\mathcal{H}$) with respect to each state variable.

We assume that $u_{1}^{}, u_{2}^{},u_{3}^{},u_{4}^{}$ are the optimal control variables for the system and $(I^{*},E_1^{*},E_2^{*},S_1^{*},S_2^{*},R_1^{*},R_2^{*})$ represents the corresponding state solutions of the system \eqref{eqnp1}.

Applying Pontryagin's Maximum Principle and taking the partial derivative of each state variable in the Hamiltonian function, we can obtain the adjoint variables $\lambda_{I}(t), \lambda_{E_1}(t), \lambda_{E_2}(t), \lambda_{S_1}(t),$ $\lambda_{S_2}(t), \lambda_{R_1}(t),$ $\lambda_{R_2}(t)$, which satisfy the canonical equations given below:

\begin{center}
\begin{eqnarray}\label{eqnqw}  
\begin{aligned}
&\frac{d\lambda_{I}(t)}{dt}=-\frac{\partial H}{\partial I}\\
&\frac{d\lambda_{E_1}(t)}{dt}=-\frac{\partial H}{\partial E_1}\\
&\frac{d\lambda_{E_2}(t)}{dt}=-\frac{\partial H}{\partial E_2}\\
&\frac{d\lambda_{S_1}(t)}{dt}=-\frac{\partial H}{\partial S_1}\\
&\frac{d\lambda_{S_2}(t)}{dt}=-\frac{\partial H}{\partial S_2}\\
&\frac{d\lambda_{R_1}(t)}{dt}=-\frac{\partial H}{\partial R_1}\\
&\frac{d\lambda_{R_2}(t)}{dt}=-\frac{\partial H}{\partial R_2}
\end{aligned}
\end{eqnarray}
\end{center}

\begin{align}
 \lambda_{I}^{'}(T)= \lambda_{E_1}^{'}(T)= \lambda_{E_2}^{'}(T)= \lambda_{S_1}^{'}(T)= \lambda_{S_2}^{'}(T)=\lambda_{R_1}^{'}(T)= \lambda_{S_2}^{'}(T)=0
\end{align}

\begin{align} \nonumber
\begin{cases}
\frac{\partial \mathcal{H}}{\partial u_{1}}=0 \hspace{.4cm} at \hspace{.4cm} u_{1}=u_{1}^{*} \\
\frac{\partial \mathcal{H}}{\partial u_{2}}=0 \hspace{.4cm} at \hspace{.4cm} u_{2}=u_{2}^{*}\\
\frac{\partial \mathcal{H}}{\partial u_{3}}=0 \hspace{.4cm} at \hspace{.4cm} u_{3}=u_{3}^{*}\\
\frac{\partial \mathcal{H}}{\partial u_{4}}=0 \hspace{.4cm} at \hspace{.4cm} u_{4}=u_{4}^{*}\\
\end{cases}
\end{align}

Thus, we obtain,

$u_{1}^{*}=\frac{(\lambda_{E_1}-\lambda_{R_1})\Tilde{E}_{1}}{b_{1}}), u_{2k}^{*}=\frac{(\lambda_{E_2}-\lambda_{R_2})\Tilde{E}_{2}}{b_{2}}),
u_{3k}^{*}=\frac{(\lambda_{S_1}-\lambda_{R_1})\Tilde{S}_{1}}{b_{3}}),
u_{4k}^{*}=\frac{(\lambda_{S_2}-\lambda_{R_2})\Tilde{S}_{2}}{b_{4}})$\\

\begin{align} 
u_{1}^{*}=
\begin{cases}
0 \hspace{2cm} if    \hspace{.4cm}    \frac{(\lambda_{E_1}-\lambda_{R_1})\Tilde{E}_{1}}{b_{1}}) < 0, \\
\frac{(\lambda_{E_1}-\lambda_{R_1})\Tilde{E}_{1}}{b_{1}})  \hspace{.4cm} if \hspace{.4cm} 0 \leq \frac{(\lambda_{E_1}-\lambda_{R_1})\Tilde{E}_{1}}{b_{1}})\leq 1,\\
1\hspace{2cm} if \hspace{.4cm}\frac{(\lambda_{E_1}-\lambda_{R_1})\Tilde{E}_{1}}{b_{1}})>1.\\
\end{cases}
\end{align}

\begin{align} 
u_{2}^{*}=
\begin{cases}
0 \hspace{2cm} if    \hspace{.4cm}    \frac{(\lambda_{E_2}-\lambda_{R_2})\Tilde{E}_{2}}{b_{2}}) < 0, \\
\frac{(\lambda_{E_2}-\lambda_{R_2})\Tilde{E}_{2}}{b_{2}})  \hspace{.4cm} if \hspace{.4cm} 0 \leq \frac{(\lambda_{E_2}-\lambda_{R_2})\Tilde{E}_{2}}{b_{2}})\leq 1,\\
1\hspace{2cm} if \hspace{.4cm}\frac{(\lambda_{E_2}-\lambda_{R_2})\Tilde{E}_{2}}{b_{2}})>1.\\
\end{cases}
\end{align}

\begin{align} 
u_{3}^{*}=
\begin{cases}
0 \hspace{2cm} if    \hspace{.4cm}   \frac{(\lambda_{S_1}-\lambda_{R_1})\Tilde{S}_{1}}{b_{3}} < 0, \\
\frac{(\lambda_{S_1}-\lambda_{R_1})\Tilde{S}_{1}}{b_{3}}\hspace{.4cm} if \hspace{.4cm} 0 \leq \frac{(\lambda_{S_1}-\lambda_{R_1})\Tilde{S}_{1}}{b_{3}}\leq 1,\\
1\hspace{2cm} if \hspace{.4cm}\frac{(\lambda_{S_1}-\lambda_{R_1})\Tilde{S}_{1}}{b_{3}}>1.\\
\end{cases}
\end{align}

\begin{align} 
u_{4}^{*}=
\begin{cases}
0 \hspace{2cm} if    \hspace{.4cm}  \frac{(\lambda_{S_2}-\lambda_{R_2})\Tilde{S}_{2}}{b_{4}}< 0, \\
\frac{(\lambda_{S_2}-\lambda_{R_2})\Tilde{S}_{2}}{b_{4}}\hspace{.4cm} if \hspace{.4cm} 0 \leq \frac{(\lambda_{S_2}-\lambda_{R_2})\Tilde{S}_{2}}{b_{4}}\leq 1,\\
1\hspace{2cm} if \hspace{.4cm} \frac{(\lambda_{S_2}-\lambda_{R_2})\Tilde{S}_{2}}{b_{4}}>1.\\
\end{cases}
\end{align}

\section{Numerical Simulation and Discussion}

In this section, we carry out simulations to investigate the effects of different parameters on the dynamics of the model (\ref{eqn1}). We choose the entering rate and the leaving rate of the population as $\xi=0.0001$ and $\delta=0.0001$, respectively and the initial values are $I(0)=1, E_1(0)=0, E_2(0)=0, S_1(0)=10/10^6, S_2(0)=1/10^6, R_1(0)=0, R_2(0)=0$.\par

\begin{figure}[H]
     \centering
     \begin{subfigure}{.48\textwidth}
         \centering
         \includegraphics[scale=0.4]{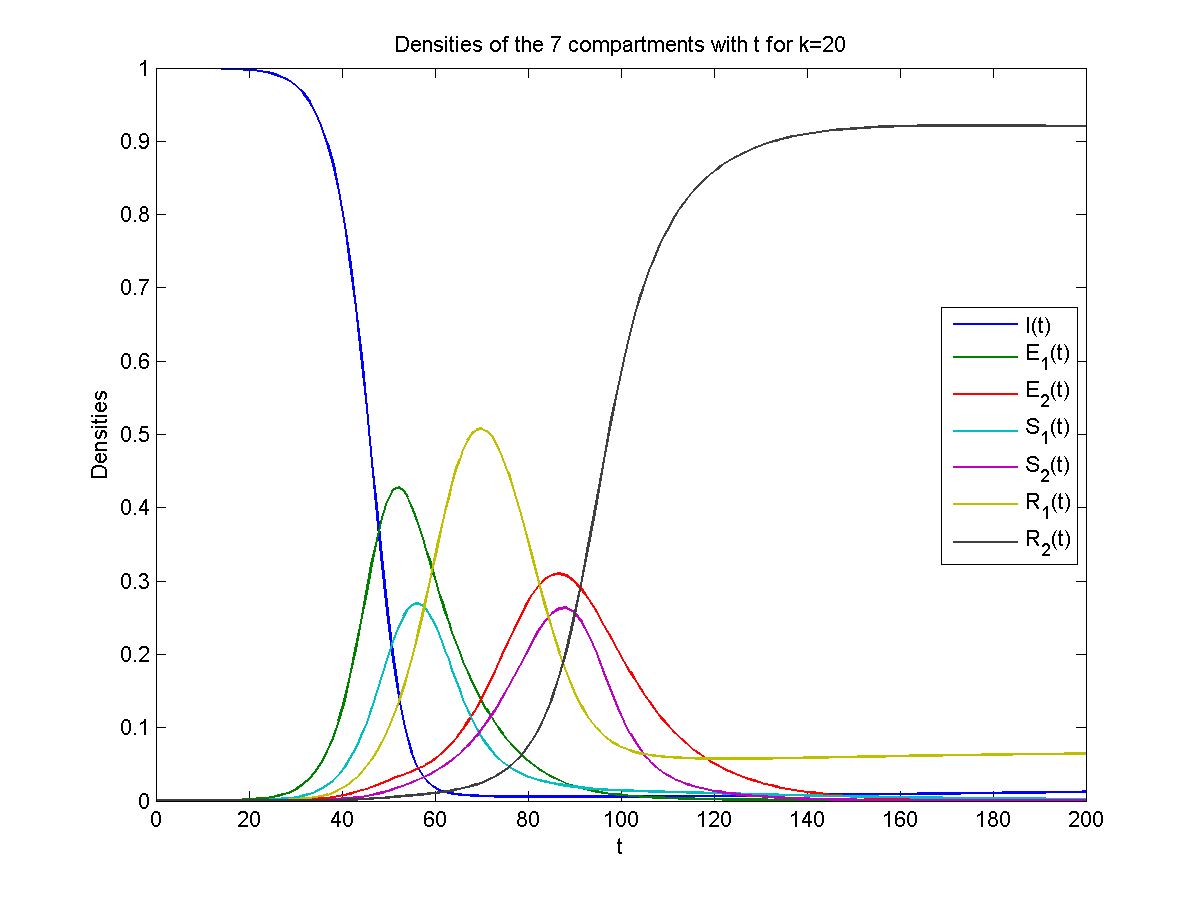}
         \caption{$\overline{k}=20$}
         \label{fig2a}
     \end{subfigure}
     \begin{subfigure}{.48\textwidth}
         \centering
         \includegraphics[scale=0.4]{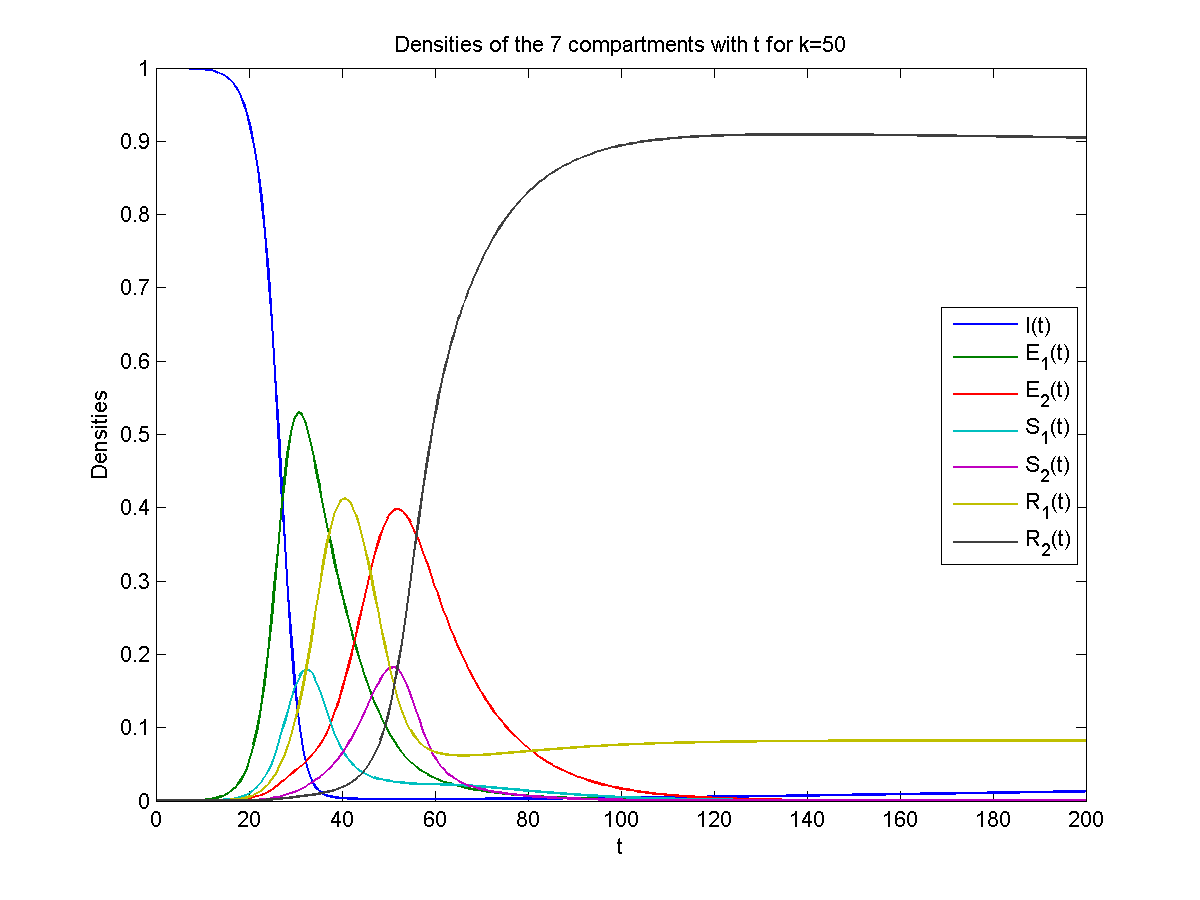}
         \caption{$\overline{k}=50$}
         \label{fig2b}
     \end{subfigure}
     \caption{The curves of the density of the seven classes over time(t)}
    \label{fig2}
\end{figure}

Fig. \ref{fig2} illustrates the change of the density of the model variables over time with average degree $\overline{k}=20$ (Fig. \ref{fig2a}) and $\overline{k}=50$ (Fig. \ref{fig2b}) of the homogeneous network. It is monitored that if the average degree of network $\overline{k}$ is increased, then the rumor spreads faster. The average degree of network accelerates the spreading of the rumor. In reality, it makes sense because if the network is connected with too many nodes, then the rumor transmits quickly through the networks.

\begin{figure}[H]
     \centering
     \begin{subfigure}{.48\textwidth}
         \centering
         \includegraphics[scale=0.4]{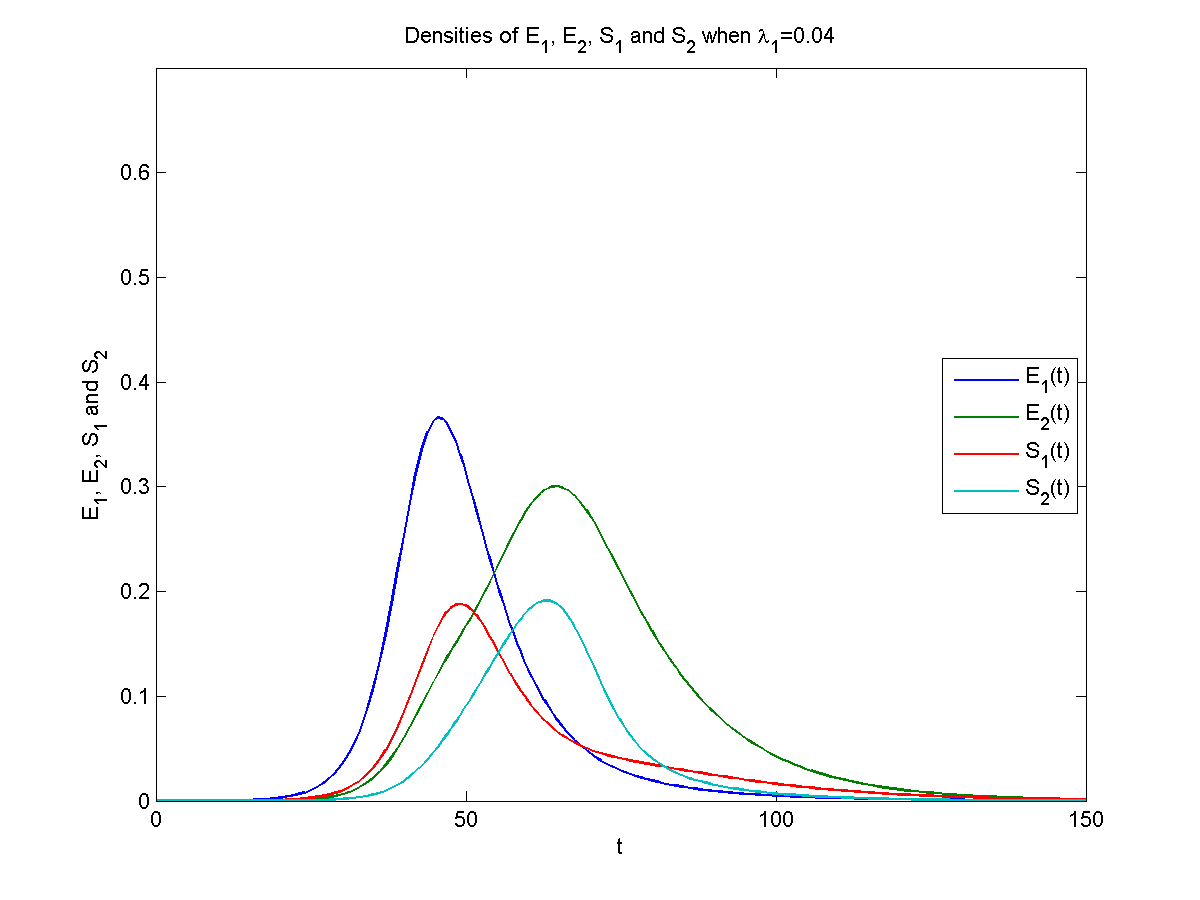}
         \caption{$\lambda_1=.04$}
         \label{fig3a}
     \end{subfigure}
     \begin{subfigure}{.48\textwidth}
         \centering
         \includegraphics[scale=0.4]{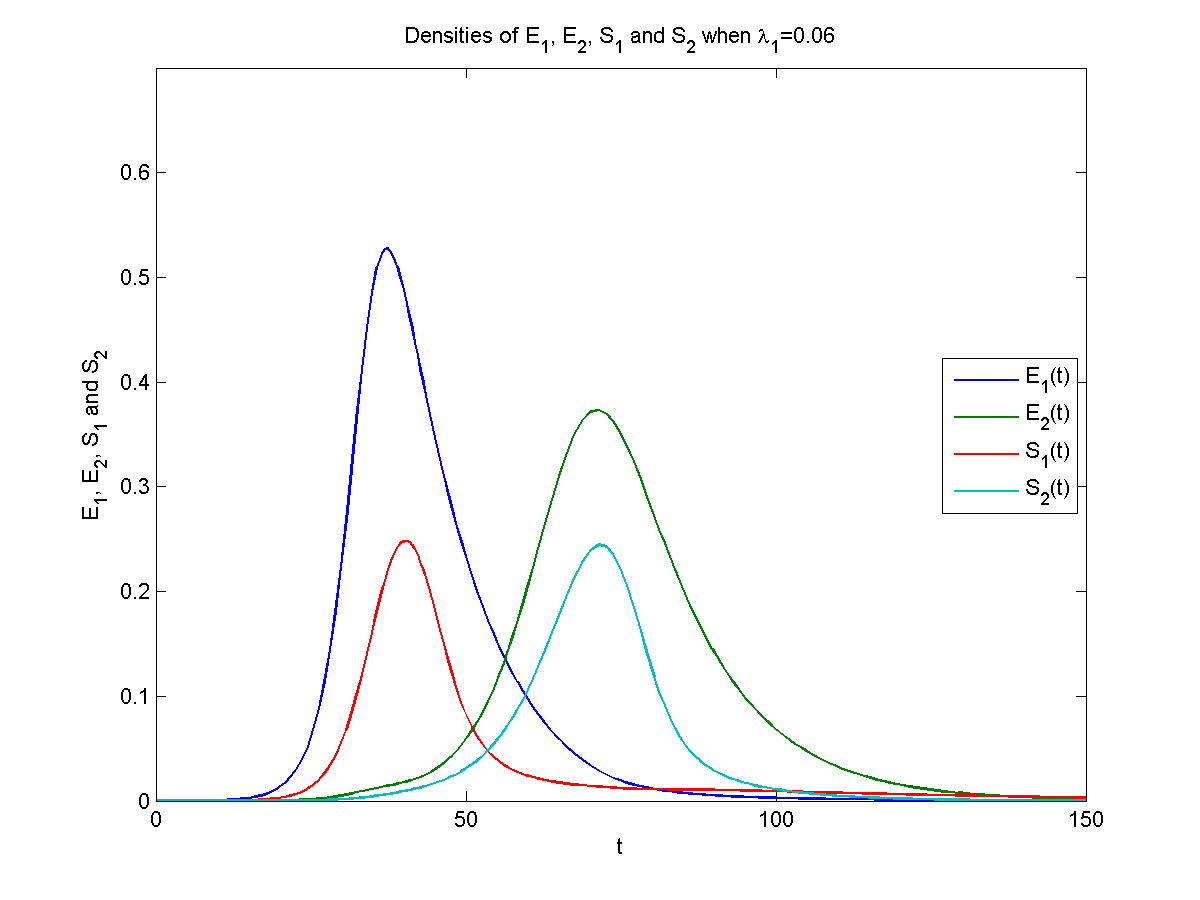}
         \caption{$\lambda_1=.06$}
         \label{fig3b}
     \end{subfigure}
     \caption{The curves of the density of $E_1, E_2, S_1$ and $S_2$ for different $\lambda_1$. }
    \label{fig3}
\end{figure}

Fig. \ref{fig3} shows the change of the density of spreaders and exposed classes for different transmission probability $\lambda_1$. It is observed that when the transmission probability $\lambda_1$ increases, the density of spreaders and exposed classes increases. In both cases, the persisting time of spreaders and exposed classes are almost the same. There is a proportional relation between the transmission probability of the ignorant to exposed and the size of the exposed as well as spreader class. To prevent the rumour from spreading, the role of this transmission probability is significant. 

\begin{figure}[H]
     \centering
     \begin{subfigure}{.48\textwidth}
         \centering
         \includegraphics[scale=0.4]{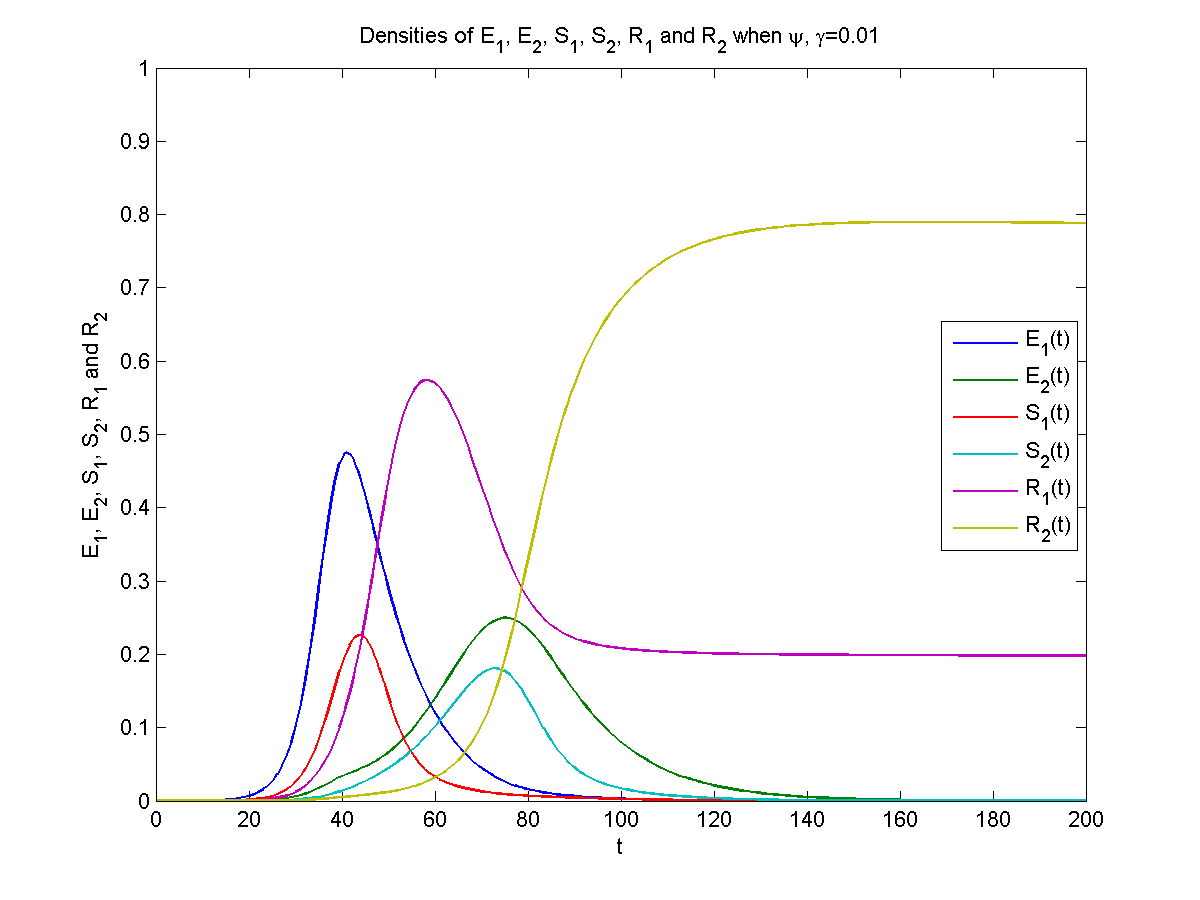}
         \caption{$\gamma, \psi = 0.01$}
         \label{fig4a}
     \end{subfigure}
     \begin{subfigure}{.48\textwidth}
         \centering
         \includegraphics[scale=0.4]{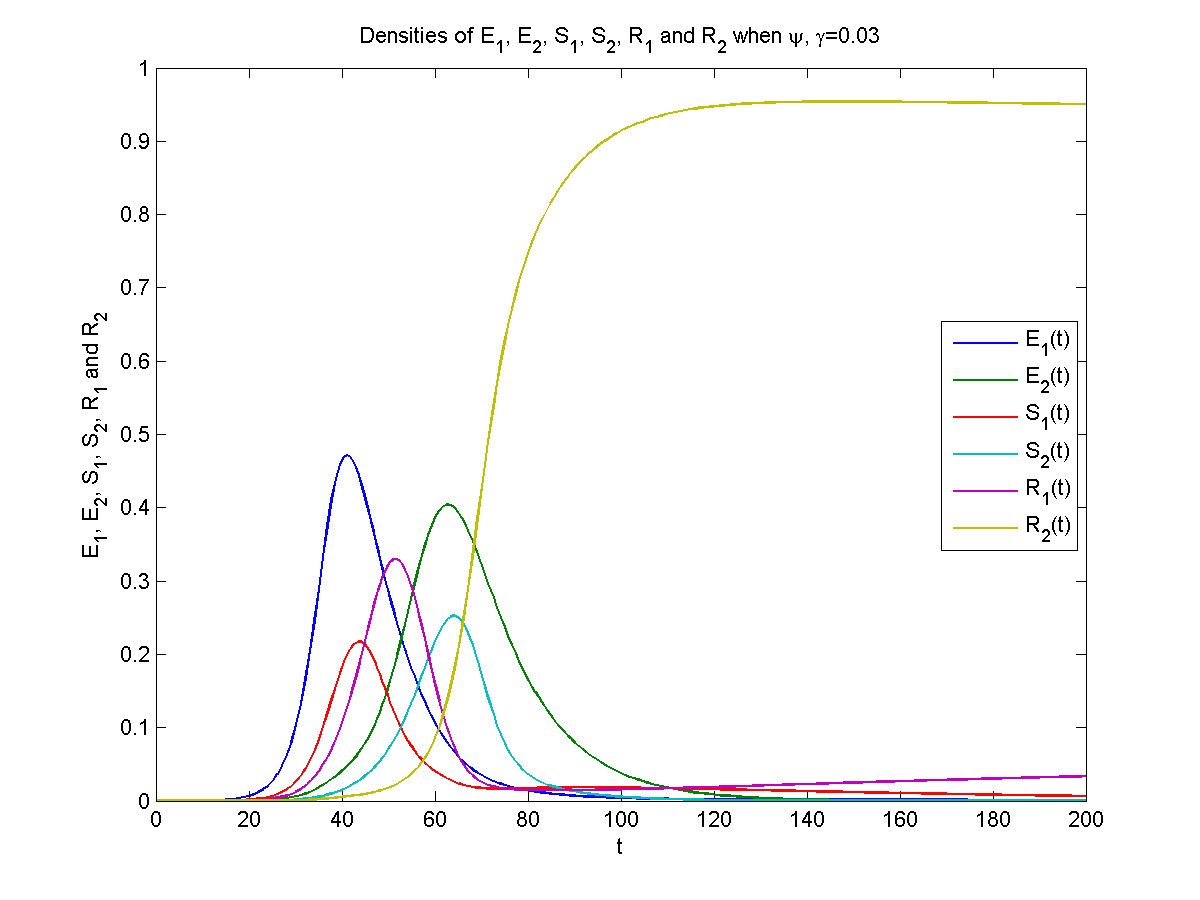}
         \caption{$\gamma, \psi = 0.03$}
         \label{fig4b}
     \end{subfigure}
     \caption{The curves of the density of $E_1, E_2, S_1, S_2, R_1$ and $R_2$ for different $\gamma$ and $\psi$.}
    \label{fig4}
\end{figure}

Fig. \ref{fig4}  depicts the change in the density of spreaders, exposed classes, and stiflers for different transfer rates $\gamma$ and $\psi$. We see that when the transfer rates $\gamma$ and $\psi$ are increased, then the density of spreader 2 and exposed 2 increase and grow faster, but the density of stifler 1 is decreased.\\

If we assume rumor 2 is the exact information of rumor 1, then we can make people conscious by spreading the exact information. People divert from spreading rumor 1, which causes the rise of the size of spreader 2 and exposed 2 classes (conscious population). This resembles a realistic scenario.

\begin{figure}[H]
     \centering
     \begin{subfigure}{.48\textwidth}
         \centering
         \includegraphics[scale=0.4]{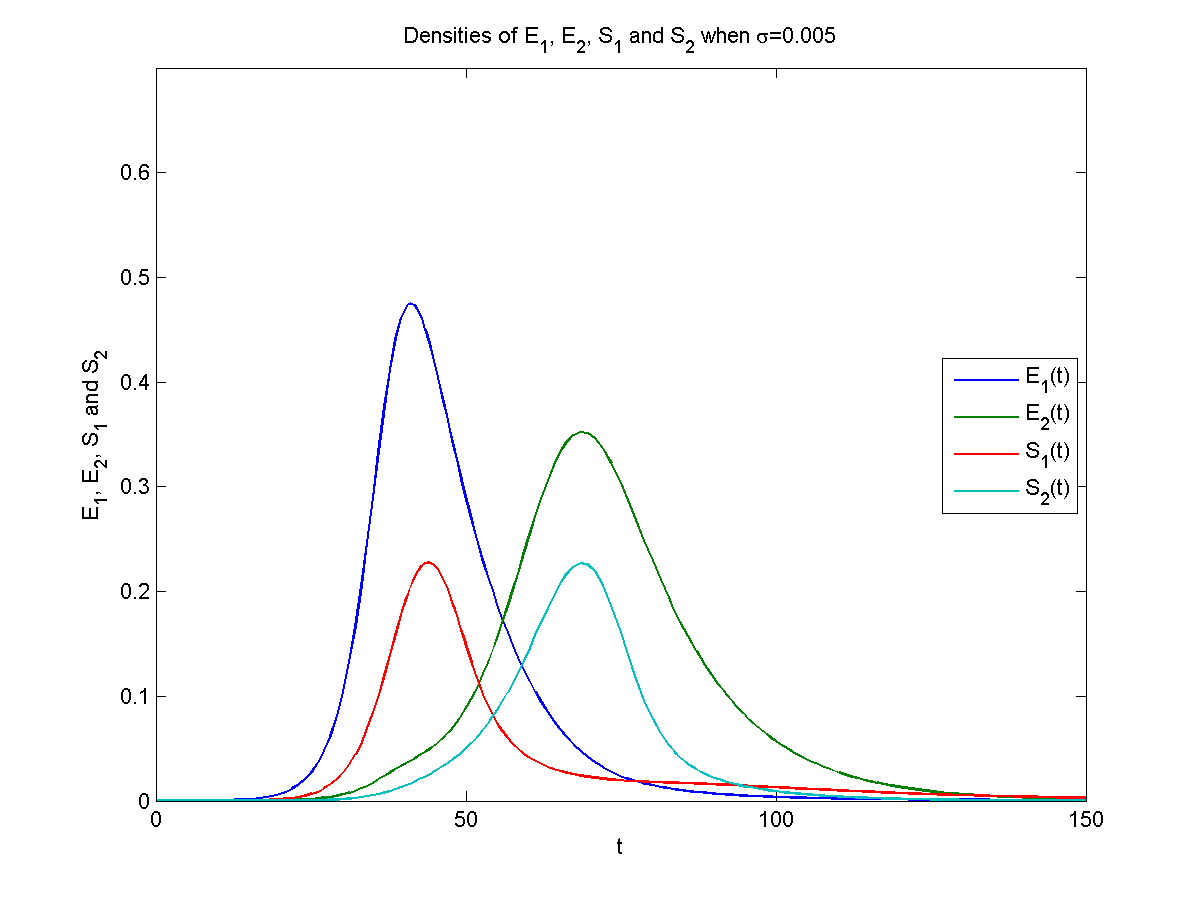}
         \caption{$\sigma=0.005$}
         \label{fig5a}
     \end{subfigure}
     \begin{subfigure}{.48\textwidth}
         \centering
         \includegraphics[scale=0.4]{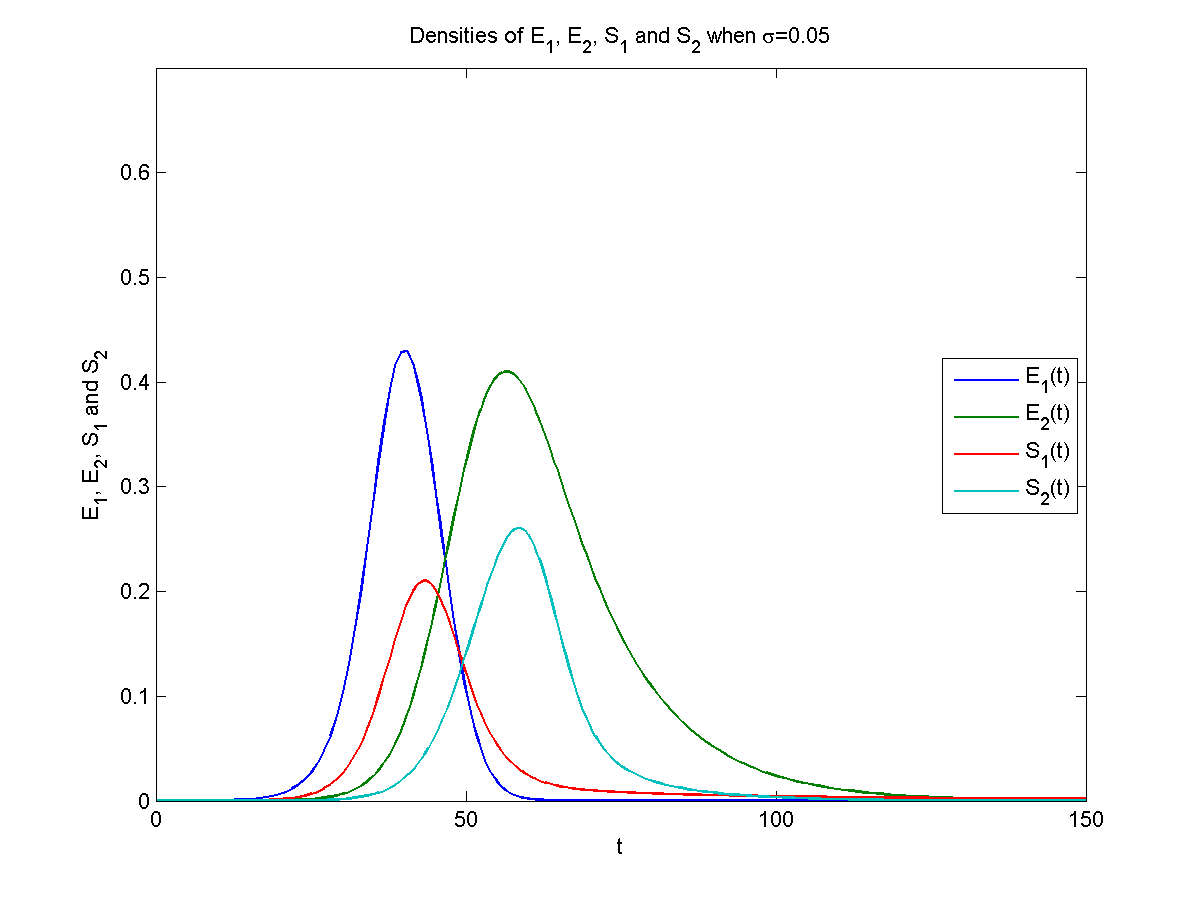}
         \caption{$\sigma=0.05$}
         \label{fig5b}
     \end{subfigure}
     \caption{The curves of the density of $E_1, E_2, S_1$ and $S_2$ over time with different $\sigma$.}
    \label{fig5}
\end{figure}

In Fig. \ref{fig5}, the effect of $\sigma$ on spreaders and exposed individuals is represented. We see that with the increase of $\sigma$, the density of spreader 2 and exposed 2 are increasing, and the density of spreader 1 and exposed 1 are decreasing.\\

If information 2 (rumor 2) is the exact information of rumor 1, then we can raise public awareness by disseminating the exact information. By raising awareness, if we can make people divert from rumor 1 to the exact information before they spread rumor 1, we can reduce the persisting duration of rumor 1  in the network. As a result, the size of the conscious population (exposed 2 and spreader 2) is increased. It is realistic that if we make people conscious of rumor 1 before spreading it by spreading the exact information, the rumor will vanish faster.

\begin{figure}[H]
     \centering
     \begin{subfigure}{.48\textwidth}
         \centering
         \includegraphics[scale=0.4]{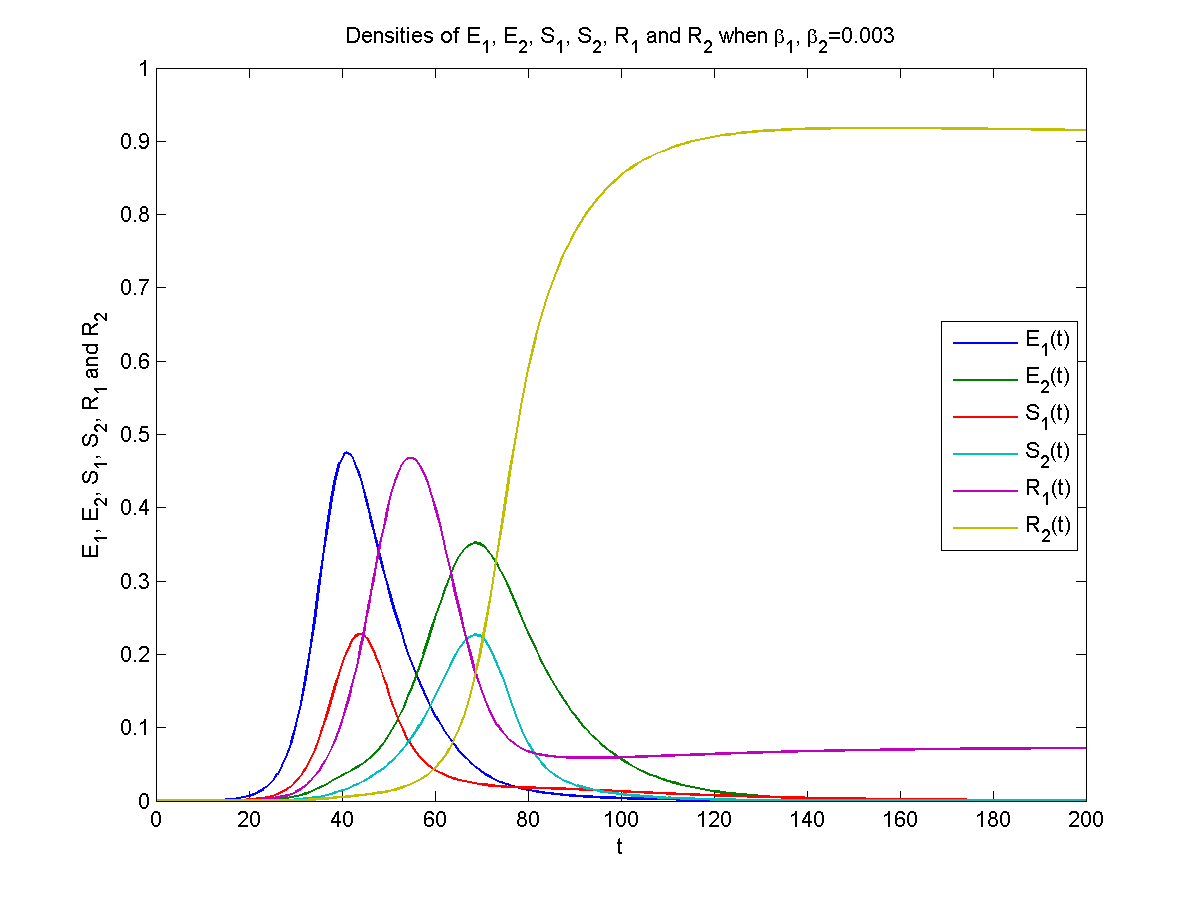}
         \caption{$\beta_1, \beta_2 = 0.003$}
         \label{fig6a}
     \end{subfigure}
     \begin{subfigure}{.48\textwidth}
         \centering
         \includegraphics[scale=0.4]{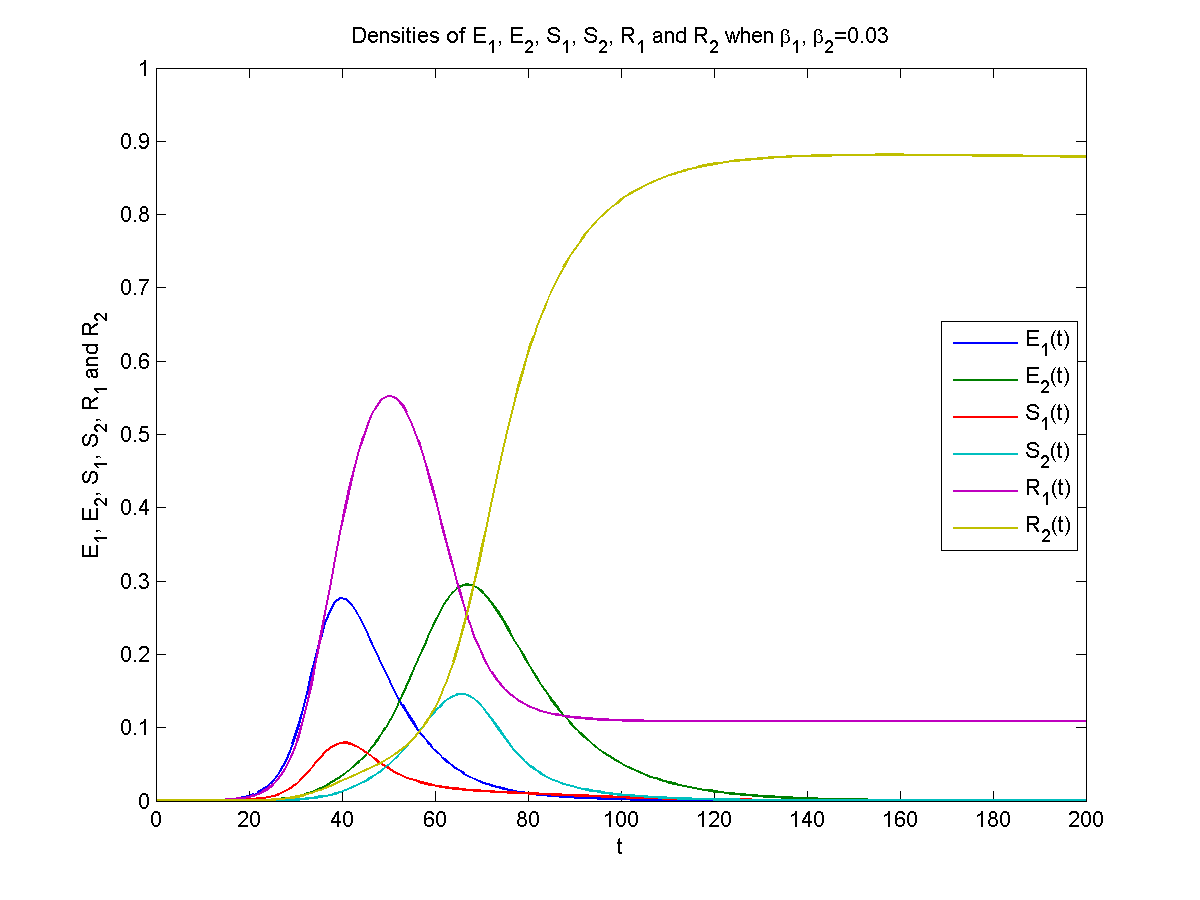}
         \caption{$\beta_1, \beta_2 = 0.03$}
         \label{fig6b}
     \end{subfigure}
     \caption{The curves of the density of $E_1, E_2, S_1, S_2, R_1$ and $R_2$ for different distrust rate $\beta_1$ and $\beta_2$.}
    \label{fig6}
\end{figure}

Fig.\ref{fig6}  depicts the effect of distrust rate $\beta_1$ and $\beta_2$ on the exposed spreaders and stiflers. The size of spreaders and exposed are inversely proportional to the distrust rate $\beta_1$ and $\beta_2$. The figure illustrates that when $\beta_1$ and $\beta_2$ increase, then the density of exposed 1, exposed 2, spreader 1, and spreader 2 decrease, and stifler 1 and stifler 2 both increase. In reality, the more the people distrust the rumor, the smaller the number of spreaders.

\begin{figure}[H]
     \centering
     \begin{subfigure}{.48\textwidth}
         \centering
         \includegraphics[scale=0.4]{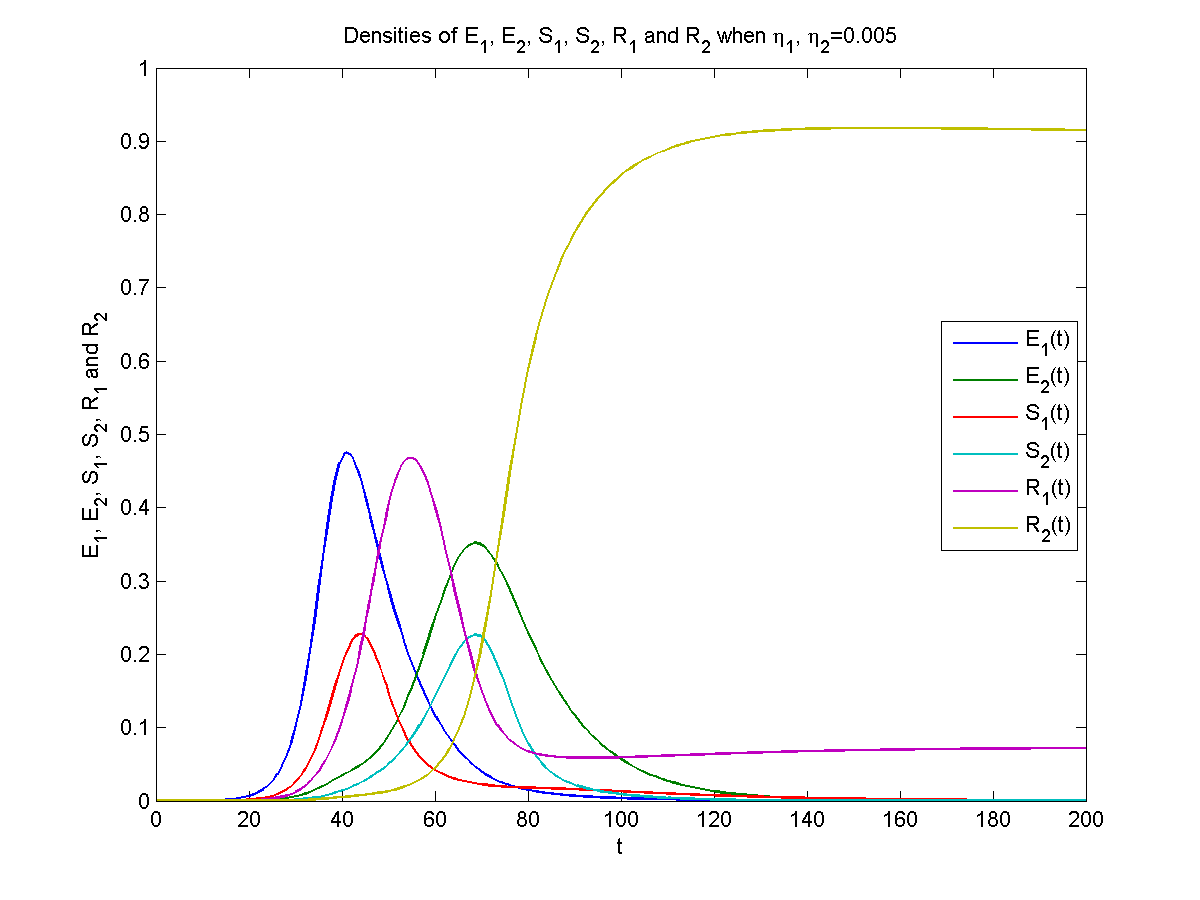}
         \caption{$\eta_1, \eta_2 = 0.005$}
         \label{fig7a}
     \end{subfigure}
     \begin{subfigure}{.48\textwidth}
         \centering
         \includegraphics[scale=0.4]{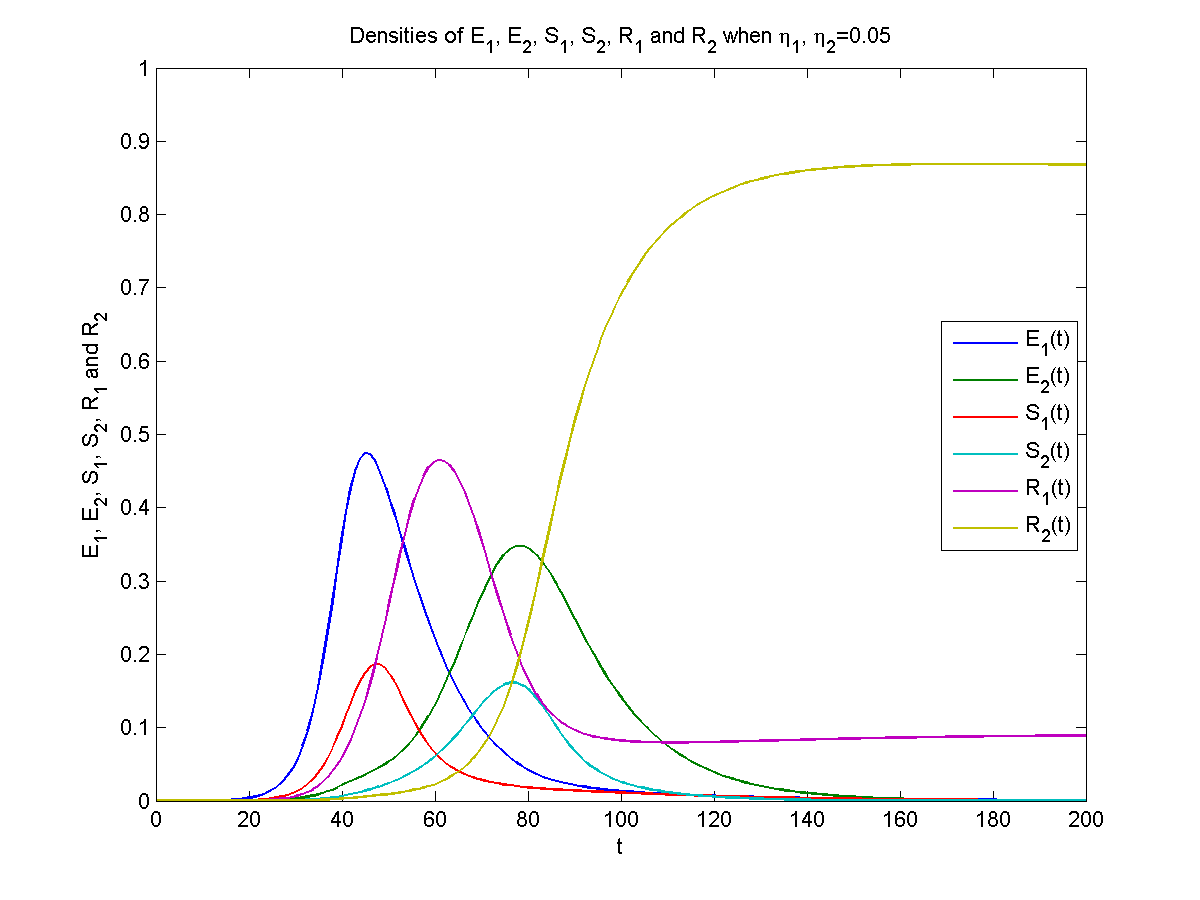}
         \caption{$\eta_1, \eta_2 = 0.05$}
         \label{fig7b}
     \end{subfigure}
     \caption{The curves of the density of $E_1, E_2, S_1, S_2, R_1$ and $R_2$ for different distrust rate $\eta_1$ and $\eta_2$.}
    \label{fig7}
\end{figure}

In Fig. \ref{fig7}, the effects of forgetting rate $\eta_1$ and $\eta_2$ on the exposed spreaders and stiflers are shown. The figure illustrates that when $\eta_1$ and $\eta_2$ increase, then the density of spreader 1 and spreader 2 decrease, and the rest are almost the same. It is realistic that when the spreaders forget the rumor, the number of spreaders will automatically decrease.

Our key observations are that the average degree of network $\overline{k}$ has a tremendously positive effect on the speed of rumor spreading. Transmission probability from ignorant to exposed 1 ($\lambda_1$) has a proportional relation with the size of all kinds of exposed and spreaders. On the other hand, diversion probabilities to exposed 2 ($\sigma$, $\psi$, and $\gamma$) have a positive effect on the size of exposed 2 and spreader 2, but only the diversion probability from exposed 1 to exposed 2 ($\sigma$) has an immense effect on the persisting duration of the exposed 1 and spreader 1. Distrust rates ($\beta_1$ for information 1 and $\beta_2$ for information 2) have inversely proportional relations with all exposed and spreader classes. We also observe that the forgetting rates ($\eta_1$ for information 1 and $\eta_2$ for information 2) have a negative effect on the spreaders. 

\subsection{ Effect of Optimal Control Strategy }
\begin{figure}[H]
     \centering
     \begin{subfigure}{.48\textwidth}
         \centering
         \includegraphics[scale=0.4]{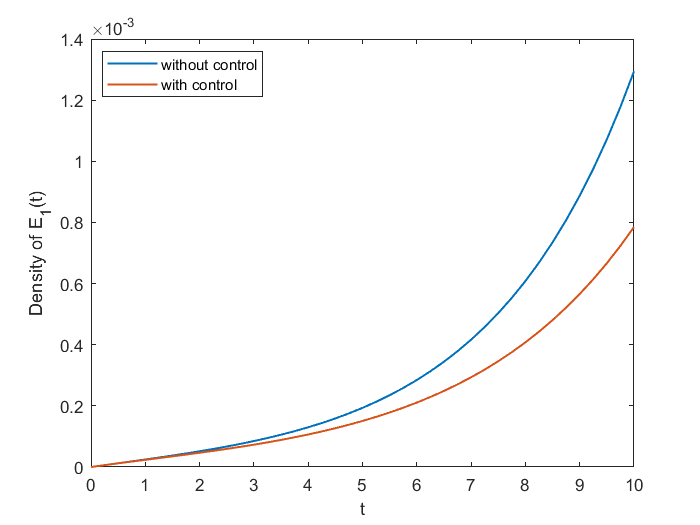}
         \caption{Densities of $E_1(t)$ over time}
         \label{fig8a}
     \end{subfigure}
     \begin{subfigure}{.48\textwidth}
         \centering
         \includegraphics[scale=0.4]{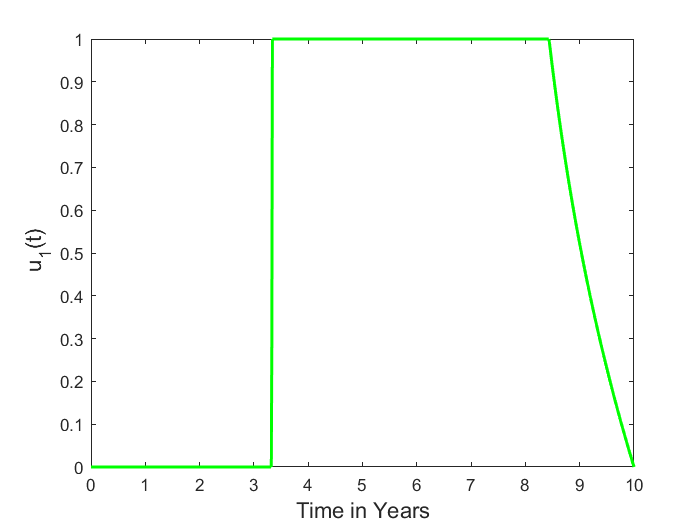}
         \caption{Effect of $u_1(t)$ over time }
         \label{fig8b}
     \end{subfigure}
     \caption{The curves of the density of $E_1 (t)$ before and after $u_1(t)$ is applied and effect of $u_1(t)$ over time }
    \label{fig8}
\end{figure}
In Fig.~\eqref{fig8a}, we observe the effect of strategy $u_1(t)$. We observe that the $E_1(t)$ density significantly decreases after strategy $u_1(t)$ is applied in the model. When authorities give official confirmation about the spread of the rumor, hesitant individuals in the exposed population in $E_1$ class become aware of the actual scenario. Hence, they do not spread the rumor anymore. Therefore, they go into the stifler class $R_1$. In Fig.~\eqref{fig8b}, we observe that the strategy is ineffective for the first few years. After a few years, the strategy becomes most effective and stays that way for many years. After that, the effect starts to lessen gradually with time. The result is very realistic. After the authority publicly clarifies the rumor, it will take some time for most individuals to hear the clarification and be affected by it. After the strategy becomes effective, it keeps reducing the population in the $E_1$ class, which is visible in Fig.\eqref{fig8a}.

\begin{figure}[H]
     \centering
     \begin{subfigure}{.48\textwidth}
         \centering
         \includegraphics[scale=0.4]{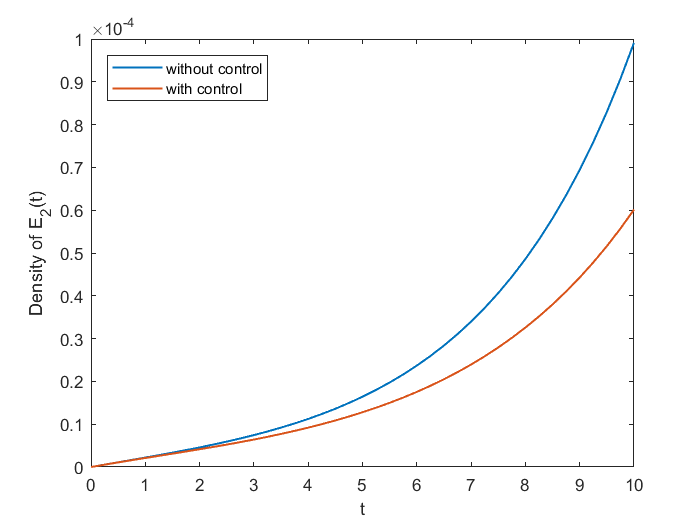}
         \caption{Densities of $E_2(t)$ over time}
         \label{fig9a}
     \end{subfigure}
     \begin{subfigure}{.48\textwidth}
         \centering
         \includegraphics[scale=0.4]{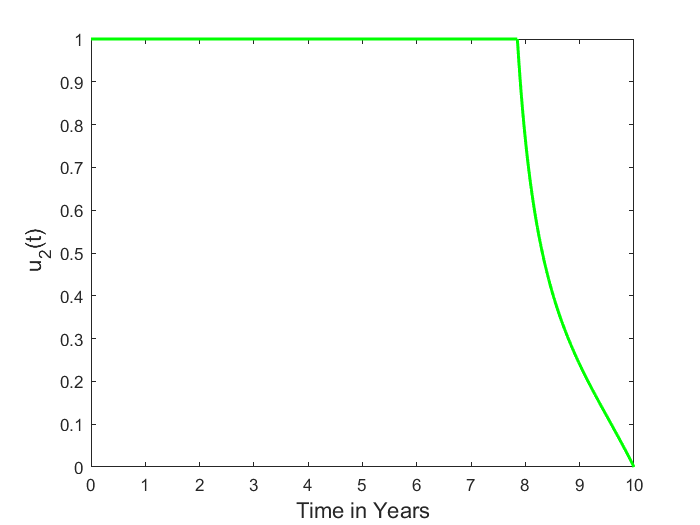}
         \caption{Effect of $u_2(t)$ over time }
         \label{fig9b}
     \end{subfigure}
     \caption{The curves of the density of $E_2 (t)$ before and after $u_2(t)$ is applied and effect of $u_2(t)$ over time }
    \label{fig9}
\end{figure}
In Fig.~\eqref{fig9a}, we observe the effect of strategy $u_2(t)$. We observe that the $E_2(t)$ density significantly decreases after strategy $u_2(t)$ is applied in the model. This is because most people in today's era use social media. Social media users are highly affected by content created by popular social media influencers. People are subconsciously highly affected when they watch didactic content on the rumors shared by their known social media influencers. Their hesitation is removed, and they start distrusting the rumor. Therefore, they do not share the rumor anymore. In Fig. \eqref{fig9b}, we observe the high influence of our control strategy $u_2(t)$ over years. It immediately starts affecting individuals in the network and stays very effective for a long time.

\begin{figure}[H]
     \centering
     \begin{subfigure}{.48\textwidth}
         \centering
         \includegraphics[scale=0.4]{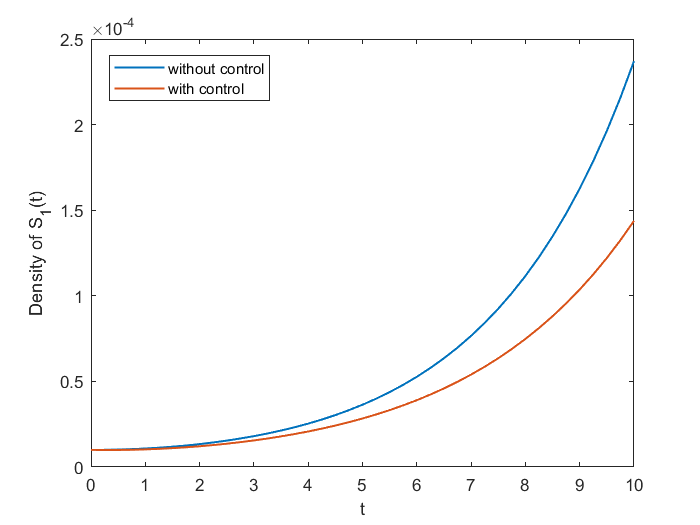}
         \caption{Densities of $S_1(t)$ over time}
         \label{fig10a}
     \end{subfigure}
     \begin{subfigure}{.48\textwidth}
         \centering
         \includegraphics[scale=0.4]{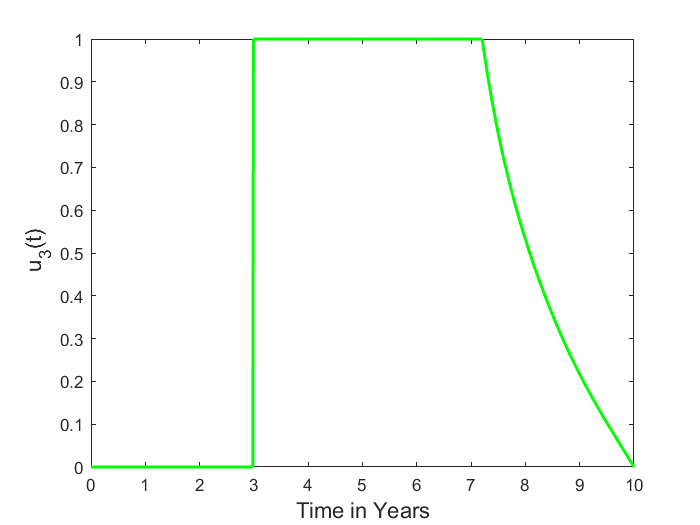}
         \caption{Effect of $u_3(t)$ over time }
         \label{fig10b}
     \end{subfigure}
     \caption{The curves of the density of $S_1 (t)$ before and after $u_3(t)$ is applied and effect of $u_3(t)$ over time }
    \label{fig10}
\end{figure}
In Fig. \eqref{fig10a}, we observe that the density of $S_1(t)$ significantly decreases after strategy $u_3(t)$ is applied in the model. The reduction happens because when the authority starts taking fines from spreaders, the spreaders become careful not to share the rumor in the network. Because, logically, nobody would want to pay a massive penalty for something as insignificant as sharing misinformation with others. So, ultimately, the spread of rumor decreases. In Fig. \eqref{fig10b}, we observe the effect of strategy $u_3(t)$. We observe that initially, it takes some time to become effective. The delay happens because when the authority starts taking fines from spreaders, the event of fining people immediately spreads only some over the social network. The process requires some time. After some time, we observe that the strategies become fully effective and stay that way for a significant time.
\begin{figure}[H]
     \centering
     \begin{subfigure}{.48\textwidth}
         \centering
         \includegraphics[scale=0.4]{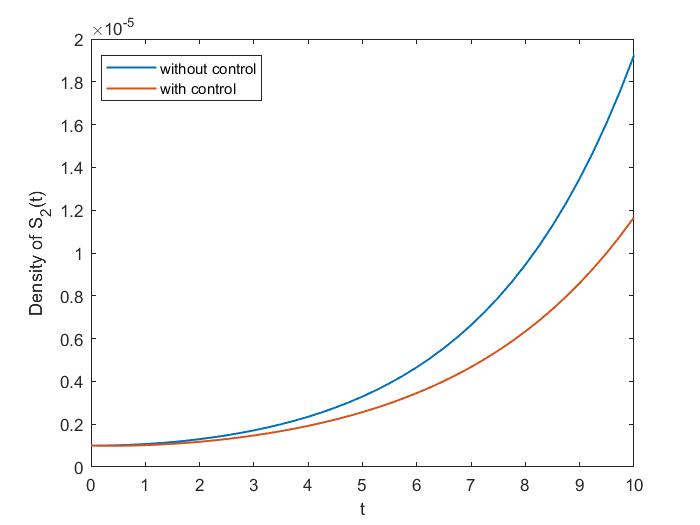}
         \caption{Densities of $S_2(t)$ over time}
         \label{fig11a}
     \end{subfigure}
     \begin{subfigure}{.48\textwidth}
         \centering
         \includegraphics[scale=0.4]{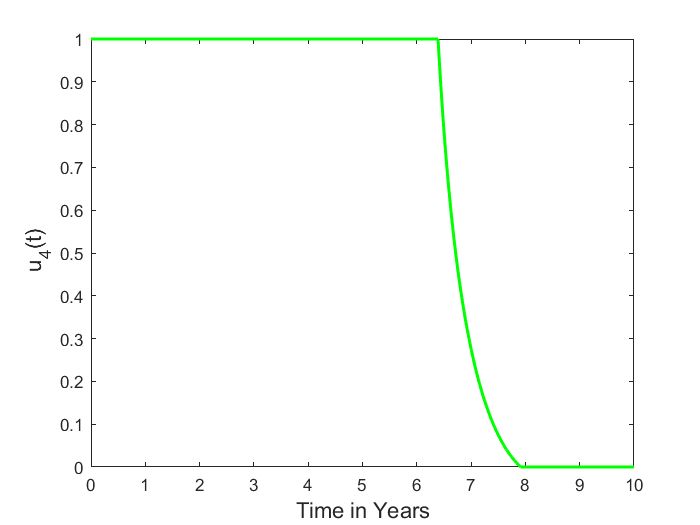}
         \caption{Effect of $u_4(t)$ over time }
         \label{fig11b}
     \end{subfigure}
     \caption{The curves of the density of $S_2 (t)$ before and after $u_4(t)$ is applied and effect of $u_4(t)$ over time }
    \label{fig11}
\end{figure}
Finally, in Fig. \eqref{fig11a}, we observe that the control strategy $u_4(t)$ effectively decreases the population density in the $S_2$ class. In Fig. \eqref{fig11b}, we observe the effectiveness of strategy $u_4(t)$ over ten years. This strategy is very effective from the initial time and remains effective for an extended period. This is because when authority starts incarcerating the spreaders (of the second rumor) and disseminating this step in the social network, people become terrified of the consequences. So they strictly abstain from sharing the rumor and enter the $R_2 $ class.

\section{Conclusion}
In this article, a new I2E2S2R rumour-spreading model in a homogeneous network has been formulated and analyzed rigorously. We have shown that the model has a rumor-free equilibrium, which is locally as well as globally asymptotically stable when the threshold parameter $\mathcal{R}_0$ is less than unity. That is, if the quantity $\mathcal{R}_0$ is less than unity, the rumor will not spread out in the network, irrespective of the initial size of the nodes. We have also shown that the rumor prevailing equilibrium of the model is globally asymptotically stable for a special case. To control the spread of rumors, we propose an optimal control problem for our proposed model with four optimal control strategies. Using Pntriagin’s maximum principle,
we study some of the mathematical properties of our proposed problem. Moreover, the connectivity of the network to many more nodes enables the rumor to circulate faster. However, the increasing transfer rate from ignorant to exposed accelerates the spreading of the rumor. We have also found that spreading the true information of the rumor may help to disappear the rumor from the network. Also, distrusting the rumor as well as forgetting it may help to reduce its spreading. Furthermore, we have found that our control strategies are very effective and realistic in controlling the spread of rumors in the network. Our strategies are most effective when applied and in the earliest period of a rumor spread.\\
\\
 {\bf Declaration of competing interest}
 The authors do not have any declaration of interest to make.
\bibliographystyle{unsrt}
\bibliography{references.bib}
\end{document}

%% file: prembles.tex
\usepackage[utf8]{inputenc}

\usepackage[table]{xcolor}
\usepackage{multirow}
\definecolor{lightgray}{gray}{0.9}
\usepackage[usestackEOL]{stackengine}
\usepackage{setspace}
\usepackage[square, numbers, comma, sort&compress]{natbib}
\usepackage{graphicx}
\graphicspath{{Figures/}}
\usepackage{tikz}
\usetikzlibrary{matrix,shapes,arrows,positioning,chains}
\usepackage[bookmarksnumbered, colorlinks,plainpages]{hyperref}
\hypersetup{colorlinks=true, linkcolor=violet, citecolor=cyan, urlcolor=teal} 
\usepackage{parskip}

\usepackage{caption}
\usepackage{subcaption}
\usepackage{color}
\usepackage{lipsum}
\usepackage{setspace}
\usepackage[english]{babel}
\usepackage{amsmath} 
\usepackage{textgreek}
\usepackage{tipa}
\usepackage{float}
\usepackage{url}
\usepackage{setspace}
\usepackage{amsfonts}
\usepackage{amssymb}
\usepackage{autobreak}
\usepackage{breqn}
\usepackage{array}
\usepackage{fancyhdr}
\usepackage{fancyhdr,ifthen}
\usepackage{blindtext}
\usepackage{mathtools}
\usepackage{amssymb}
\usepackage{pdflscape}
\usepackage{tocloft} 

\emergencystretch=\maxdimen